\documentclass[letter,11pt]{article}
\pdfoutput=1 

\usepackage{jcappub} 

\usepackage[T1]{fontenc} 

\usepackage{amssymb}
\usepackage{graphicx, epsfig, bm, amsmath}
\usepackage{hyperref}

\usepackage{ mathrsfs }
  \usepackage{dsfont}

\definecolor{darkgreen}{cmyk}{0.85,0.2,1.00,0.2}

\usepackage{color}
\usepackage{ifthen}

\def\nn{\nonumber}
\def\({\left(}
\def\){\right)}
\def\[{\left[}
\def\]{\right]}

\def\f{\chi}
\newcommand{\beq}{\begin{equation}}
\newcommand{\beqn}{\begin{eqnarray}}
\newcommand{\eeq}{\end{equation}}
\newcommand{\eeqn}{\end{eqnarray}}

\bibstyle{JHEP}

\title{Fermion production during and after axion inflation}

\author{Peter Adshead and}
\author{Evangelos I. Sfakianakis}

\affiliation{Department of Physics, University of Illinois at Urbana-Champaign, Urbana, Illinois 61801, U.S.A.}
\emailAdd{adshead@illinois.edu}

\emailAdd{esfaki@illinois.edu}

\abstract{We study derivatively coupled fermions in axion-driven inflation, specifically $m_\phi^2\phi^2$ and monodromy inflation, and calculate particle production during the inflationary epoch and the post-inflationary axion oscillations. During inflation, the rolling axion acts as an effective chemical potential for helicity which biases the gravitational production of one fermion helicity over the other. This mechanism allows for efficient gravitational production of heavy fermion states that would otherwise be highly suppressed. Following inflation, the axion oscillates and fermions with both helicities are produced as the effective frequency of the fermion field changes non-adiabatically. For certain values of the fermion mass and axion-fermion coupling strength, the two helicity states are produced asymmetrically, resulting in unequal number-densities of left- and right-helicity fermions.
}

\begin{document}
\maketitle
\flushbottom

\section{Introduction}
\label{sec:intro}

In this work we study the behavior of fermionic degrees of freedom during and immediately following inflation driven by an axionic field \cite{Freese:1990rb, Adams:1992bn} (for a recent review of axion inflation see \cite{Pajer:2013fsa} and references therein). Specifically, we consider Majorana fermions coupled derivatively to the inflaton field via a dimension-five operator.  When the fermions have degenerate masses and are paired to make Dirac fermions, this is the coupling to the axial-vector current, a conserved current in the massless limit (in the absence of anomalous gauge couplings).  

In local thermal equilibrium, this coupling between the fermion fields and the rolling axion acts as a chemical potential for helicity, allowing the system to reduce its free energy by converting some fermions of one helicity into the other. The universe is not in thermal equilibrium during inflation, however, in this work we demonstrate that this coupling has a similar effect on the gravitational production of fermions. During inflation, the exponential expansion of spacetime produces particles from the vacuum as their wavelength becomes comparable to the Hubble length. In the absence of the axion coupling, all spin, or helicity states are populated with a probability that depends only on the fermion mass, $m_{\psi}$, and the Hubble rate, $H$. This gravitational pair production is most efficient for small masses, with the production rate for larger masses being exponentially suppressed by the ratio $m_\psi/H$. In the limit that the fermions are massless, no particle production occurs because the theory is conformally equivalent to a Minkowski space theory (see e.g.\ \cite{Parker:2009uva}). We show that the coupling to the rolling axion leads to the biasing of this gravitational fermion-production, which enhances the production rate of one of the helicities while suppressing the other. Furthermore, this enhancement can partially counter the exponential suppression of heavy masses allowing for efficient production of fermion states that would otherwise be heavily suppressed. On the other hand, in the limit that the fermion mass vanishes, the axion-fermion coupling reduces to a boundary term which has no effect on the equations of motion and the effect vanishes. 

The significant feature arising from considering a pseudo-scalar axion field as the inflaton, that we will discuss here, is the nature of the derivative coupling to fermions. The specific form of the axion potential appears to play a role in the details, but does not change the mechanism and the qualitative results of the paper. It  is shown that even the final results only differ by an ${\cal O}(1)$ factor for a significant parameter range in two different forms of the inflaton potential we considered.

Derivative couplings of vector gauge-field degrees of freedom to rolling pseudo-scalar fields \cite{Garretson:1992vt, Prokopec:2001nc,Barnaby:2011vw, Barnaby:2011qe, Shiraishi:2013kxa, Cook:2013xea, Adshead:2015pva} lead to strong growth of one of the polarizations of the gauge fields during de Sitter expansion, but the effect of such couplings on fermion fields has been little studied. We demonstrate that in the fermionic case there is a similar effect, however, Pauli blocking prevents the occupation numbers from exceeding unity and consequently the growth of the fermion field is constrained.

Following the inflationary epoch, the axion oscillates about the minima of its potential which causes the effective frequency of the fermion fields to vary non-adiabatically. These oscillations cause both helicities to be produced in turn as the effective momentum of the left- and right-helicity fermions vanishes. However, the axion oscillations damp as the universe expands which reduces the efficiency of successive particle-production events. In combination with the inflationary production, this damping can lead to asymmetric number densities of fermions with left and right helicities. Moreover,  since neutrino helicity is equal to lepton number in the standard model, this effect may be relevant for the generation of the matter-antimatter asymmetry in the universe - we pursue this question in a companion paper \cite{Adshead:2015jza}.

It is well established that fermions can be produced via Yukawa couplings to oscillating scalar fields after inflation  \cite{Greene:1998nh, Greene:2000ew, Peloso:2000hy,GarciaBellido:2000dc,Tsujikawa:2000ik}. Preheating in these models is, in some instances,  known to produce fermions with masses much larger than the inflaton mass and may be important for leptogenesis scenarios where the baryon asymmetry is produced by the decay of  right-handed neutrinos with masses near the grand unified energy (GUT) scale \cite{Giudice:1999fb, Barbieri:1999ma}.  Further, Ref.\ \cite{Pearce:2015nga}, studied  neutrino production during the relaxation and oscillations of the Higgs condensate following inflation as a mechanism for the generation of a matter-antimatter asymmetry via leptogenesis. The cosmological consequences of the gravitational production of heavy fermionic states during inflation is well appreciated \cite{Lyth:1996yj}, and gravitational production has been considered as a mechanism for producing super-heavy fermions during inflation  \cite{Chung:2011ck, Chung:2013rda}. The behavior of fermions coupled to axions in the post-inflationary universe was first considered in the context of spontaneous baryogenesis \cite{Cohen:1987vi, Cohen:1988kt, Turner:1988sq, Dolgov:1996qq, Dolgov:1994zq} and has more recently been revived for leptogenesis scenarios \cite{Yamaguchi:2002vw, Kusenko:2014uta}.  Our work on fermion production differs from these existing works because we consider both the production during the inflationary period due to the axion coupling, which has significantly different phenomenology, as well production following inflation during the reheating phase.

We work in natural units where $c = \hbar = 1$ and denote the reduced Planck mass by $M^{-1}_{\rm Pl} = \sqrt{8\pi G}$.

\section{Background and conventions}

Our notation and conventions for the axion, gravitational, and fermion sectors of the theory are as follows.
We consider an axion inflationary sector minimally coupled to Einstein gravity
\begin{align}
S_{\rm inf} = \int {\rm d}^4 x \sqrt{-g}\Big[\frac{M_{\rm Pl}^2}{2}R - \frac{1}{2}(\partial\phi)^2 - & V(\phi)\Big],
\end{align}
where $V(\phi)$ is a potential capable of supporting an extended period of slow-roll inflation. For concreteness we consider the simplest chaotic inflation model with a quadratic potential \cite{Linde:1983gd, Linde:1984st}
\begin{align}
V(\phi) = {1\over 2}m_\phi^2 \phi^2 \, ,
\end{align}
as well as the potential from the simplest type of axion-monodromy inflation \cite{Silverstein:2008sg, McAllister:2008hb}
\begin{align}\label{eqn:monopot}
V(\phi) = \mu^3 (\sqrt{\phi^2 + \phi_c^2} - \phi_c).
\end{align}
In addition to the inflaton sector, we consider a set of fermions\footnote{We work with two-component spinors and consider only fields that transform under the left-handed representation of the Lorentz group. Right-handed fields can be obtained simply by complex conjugation, that is, if $\chi_i$ is a left-handed field, $\chi_i^\dagger$ is right-handed. See Appendix \ref{app:notation} for some key notations used in this work. Refer to \cite{Dreiner:2008tw} for an extensive review.} $\tilde{\f}_i$, described by the action
 \begin{align}
S_{\rm f} = \int {\rm d}^4 x \sqrt{-g}\Big[ i\tilde\f_i^\dagger{}_{\dot\alpha} e^{\mu}{}_{a}\bar{\sigma}^{a \dot\alpha \beta}D_{\mu} \tilde\f_{i}{}_{\beta} &-\frac{1}{2}m_{ij} (\tilde\f_i^{\alpha}\tilde \f_j{}_{\alpha} +\tilde\f_i^\dagger{}_{\dot\alpha} \tilde\f^{\dagger}_{j}{}^{\dot\alpha}) +\frac{C_{i}}{f}\partial_{\mu}\phi\tilde\f_i^\dagger{}_{\dot\alpha} e^{\mu}{}_{a}\bar{\sigma}^{a \dot\alpha \beta}\tilde\f_{j}{}_{\beta}\Big],
\end{align}
where $m_{ij}$ is hermitian, $C_{i}$ is real, and $f$ is a mass scale associated with the axion. The covariant derivative of the spinor fields is
\begin{align}
D_{\mu} = \partial_{\mu}+\frac{1}{8}\omega_{\mu a b}\[\sigma^{a}, \bar{\sigma}^b\],
\end{align}
where the spin connection is
\begin{align}
\omega_{\mu a b } = & \frac{1}{2}e_{a}{}^{\nu}(\partial_{\mu}e_{b\nu} - \partial_{\nu}e_{b\mu}) 
- \frac{1}{2}e_{b}{}^{\nu}(\partial_{\mu}e_{a\nu} - \partial_{\nu}e_{a\mu}) 
-\frac{1}{2}e_{a}{}^\nu e_{b}{}^{\lambda}(\partial_\nu e_{c \lambda} - \partial_{\lambda}e_{c\nu})e^c{}_{\mu}.
\end{align}
We work in cosmic time and ignore metric fluctuations as well as inflaton fluctuations. While fluctuations of the axion and gravitational fields are likely to be interesting, our primary interest in this work is the behavior of the fermion fields in the classical, homogeneous axion and gravitational backgrounds. We work in `mostly plus' convention for the metric with
\begin{align}
ds^2 = -dt^2 + a^2 d{\bf x}^2 = e^a{}_{\mu}e^b{}_{\nu}\eta_{ab},
\end{align}
where $a(t)$ is the scale factor and $e^a{}_{\mu}$ are vierbeins.

We rescale our fermion fields by $a^{-3/2}$ to absorb the factor of $\sqrt{-g}$ in the action and turn the covariant derivative into a partial derivative. We write $\f_i = a^{3/2}\tilde{\f}_i$ so that the fermion action reads
 \begin{align}\label{eqn:fermionaction}
S_{\rm f} =  \int {\rm d}^4 x \Big[ i\f_i^\dagger{}_{\dot\alpha} e^{\mu}{}_{a}\bar{\sigma}^{a \dot\alpha \beta}\partial_{\mu} \f_{i}{}_{\beta} &-\frac{1}{2}m_{ij} (\f_i^{\alpha} \f_j{}_{\alpha} +\f_i^\dagger{}_{\dot\alpha} \f^{\dagger}_{j}{}^{\dot\alpha}) +\frac{C_{i}}{f}\partial_{\mu}\phi \f_i^\dagger{}_{\dot\alpha} e^{\mu}{}_{a}\bar{\sigma}^{a \dot\alpha \beta}\f_{i}{}_{\beta}\Big].
\end{align}
For simplicity, we consider only a pair of fermions, $i \in \{1, 2\}$.  Further, we treat separately the case where the mass matrix $m_{ij}$ has degenerate eigenvalues, in which case there is an $O(2)$ flavor symmetry and the fields can be paired to make a Dirac fermion. 

Above and in what follows, Greek letters from the middle of the alphabet  represent four-dimensional space-time indices, while Greek letters from the beginning of the alphabet denote spinor indices. Roman letters from the start of the alphabet denote four-dimensional Lorentz indices, while Roman indices from the middle of the alphabet  denote spatial indices as well as flavor indices. Einstein summation convention is used for spacetime, Lorentz and repeated flavor indices, unless otherwise noted. Paired ascending dotted indices and descending undotted spinor indices will also be summed.  

\section{Majorana and Dirac fermions}

There are subtle differences between the cases where the fermions can be combined into a single Dirac fermion and can carry charged currents compared with the case where they are Majorana fermions. We treat these cases separately in what follows and then consider their vacuum states and quasi-particle numbers in expanding Friedmann-Robertson-Walker spacetimes.

\subsection{Non-degenerate fermions: Majorana fermions}

First we consider the case where the fermions have non-degenerate masses and there are no conserved currents. Without loss of generality, we work in a basis where $m_{ij}$ is diagonal, and the fermions have masses $m_i$. Variation of the action at Eqn.\ \eqref{eqn:fermionaction} with respect to $\f_i^\dagger{}_{\dot\alpha}$ yields the equations of motion for these Majorana fermions
\begin{align}\label{eqn:chieom}
 i e^{\mu}{}_{a} \bar{\sigma}^{a \dot{\alpha}\beta}\partial_{\mu}\chi_{i \beta}+\frac{C}{f}\partial_{\mu}\phi e^{\mu}{}_{a}  \bar{\sigma}^{a \dot\alpha \beta} \chi_{i}{}_{\beta} = m_i \chi_i^{\dagger\dot\alpha}.
\end{align}
It is convenient to expand each field into a Fourier basis,
\begin{align}\label{eqn:chimode}
\chi_{i\alpha}({\bf x}, t) = \sum_{\lambda  = \pm 1}\int \frac{{\rm d}^3 k}{(2\pi)^3}\[ x^\lambda _{i\alpha}({\bf k}, t)a_{i\bf k}^\lambda  e^{i{\bf k}\cdot{\bf x}}+ y^\lambda _{i\alpha}({\bf k}, t)a_i^\dagger{}^\lambda _{\bf k} e^{-i{\bf k}\cdot{\bf x}} \],
\end{align}
where we have introduced  creation and annihilation operators, $a_{i \bf k}^\lambda$ and $a_{i \bf k}^{\dagger\lambda}$, which  satisfy the  anti-commutation relations
\begin{align}
\{a^\lambda _{i\bf k}, a^{\dagger \lambda '}_{j\bf k'}\} = (2\pi)^3\delta^3({\bf k}- {\bf k}')\delta_{\lambda \lambda '}\delta_{ij},
\end{align}
with all other anti-commutators vanishing, as usual. Here and throughout we use $\lambda = \pm 1$ to denote the spin or helicity states of the fermions. It follows that
\begin{align}\label{eqn:chidaggermode}
\chi^\dagger_{i\dot\alpha}({\bf x}, t) = \sum_{\lambda = \pm 1} \int \frac{{\rm d}^3 k}{(2\pi)^3}\[ x^{\lambda  \dagger}_{i\dot\alpha}({\bf k}, t)a^\dagger{}^\lambda _{i \bf k}  e^{-i{\bf k}\cdot{\bf x}}+ y^{\lambda \dagger}_{i\dot\alpha}({\bf k}, t)a_{i\bf k}^\lambda  e^{i{\bf k}\cdot{\bf x}} \].
\end{align}
We quantize the fields by imposing the anti-commutation relations
\begin{align}
\{\chi_{i\alpha}({\bf x}, t), \pi^\beta_{\chi_j} ({\bf y}, t) \} = i\delta^{\beta}_{\alpha} \delta^{3}({\bf x} - {\bf y})\delta_{ij},
\end{align}
where the canonical momenta of the fermion fields are found in the usual way
\begin{align}
\pi^\beta_{\chi_i}  = \frac{\partial\mathcal{L}_{\rm f}}{\partial \dot\chi_{i \beta}}= & i \chi_i^\dagger{}_{\dot\alpha} \bar{\sigma}^{0\dot\alpha \beta} ,
\end{align}
where here and throughout an overdot on a field denotes a derivative with respect to cosmic time, i.e. $\dot\chi_{i \beta} = \partial_t \chi_{i \beta}$. 
After inserting the mode expansions, the fields are canonically quantized provided that  the fermion wavefunctions are normalized according to (no sum on $i$)
\begin{align}
\sum_\lambda\[ x^\lambda _{i\alpha}({\bf k}, t)x^{\lambda  \dagger}_{i\dot\alpha}({\bf k}, t)\bar{\sigma}^{0\dot\alpha \beta}   +y^\lambda _{i\alpha}({\bf k}, t)y^{\lambda \dagger}_{i\dot\alpha}({\bf k}, t) \bar{\sigma}^{0\dot\alpha \beta}\] = \delta^\beta_{\alpha}.
\end{align}
Substituting the mode expansions, Eqn.\ \eqref{eqn:chimode} and \eqref{eqn:chidaggermode}, into the equation of motion, Eqn.\ \eqref{eqn:chieom}, we find the coupled equations for $x_\beta$ and $y^{\dagger \dot\beta}$
\begin{align}\label{eqn:majeqn1}
 i \(\bar{\sigma}^{0 \dot{\alpha}\beta}\partial_{0}+ i \frac{k_j}{a} \bar{\sigma}^{j \dot{\alpha}\beta}\)x^\lambda _{i \beta}({\bf k}, t)+\frac{C}{f}\dot\phi \bar{\sigma}^{0 \dot\alpha \beta} x^\lambda _{i}{}_{\beta} ({\bf k}, t) = & m_i y_i^{\lambda \dagger\dot\alpha}({\bf k}, t), \\\label{eqn:majeqn2}
 i \({\sigma}^{0}_{ \delta \dot\gamma}\partial_{0} + i \frac{k_j}{a} {\sigma}^{j}_{ \delta\dot\gamma}\)y^{\lambda  \dagger \dot\gamma}_{i}({\bf k}, t)-\frac{C}{f}\dot\phi {\sigma}^{0}_{ \delta\dot\gamma} y^{\lambda  \dagger \dot\gamma}_{i}({\bf k}, t)= & m_i x^\lambda _{l \delta}({\bf k}, t).
\end{align}
We work in a basis of helicity eigenspinors, which satisfy
\begin{align}
\vec{\sigma}\cdot\hat{k} \xi_\lambda = \lambda \xi_\lambda, \quad \lambda = \pm 1, \quad \xi_{-\lambda}(-\hat{k}) = & \iota^{\lambda}_{\hat{k}}\xi_{\lambda}(\hat{k}),
\end{align}
where $\iota^{\lambda}_{\hat{k}}$ is a phase that satisfies
\begin{align}
\iota^{\lambda *}_{\hat{k}}\iota^{\lambda}_{\hat{k}} = 1, \quad \iota^{\lambda}_{-\hat{k}} = -\iota^{\lambda}_{\hat{k}}.
\end{align}
Writing the spinors as
\begin{align}
x_{i \delta}^{\lambda}({\bf k}, t) = & X^{\lambda}_{i k} (t) \xi_\lambda({\bf k}), \quad
y^{\lambda \dagger \dot\alpha }_{i}({\bf k}, t) =  Y^{\lambda *}_{i k}(t) \xi_\lambda({\bf k}),
\end{align}
canonical quantization requires that (no sum on $i$)
\begin{align}
\sum_\lambda\[ X^{\lambda}_{i k} (t) X^{\lambda *}_{i k} (t) + Y^{\lambda *}_{i k}(t) Y^{\lambda}_{i k}(t)  \] = 1.
\end{align}
In terms of these wavefunctions, the equations of motion \eqref{eqn:majeqn1} and \eqref{eqn:majeqn2} become
 \begin{align}\label{eqn:eomsmaj1st}
i \(\partial_{t}- i \frac{k}{a} \lambda \)X^{\lambda}_{i k} (t) +\frac{C}{f}\dot\phi X^{\lambda}_{i k} (t) = & m_i Y^{\lambda *}_{i k}(t) ,
\\
 i \(\partial_{t} + i  \frac{k}{a}\lambda  \)Y^{\lambda*}_{i k}(t)-\frac{C}{f}\dot\phi Y^{\lambda *}_{i k}(t) = & m_i X^{\lambda}_{i k} (t).
\end{align}
These coupled equations can be decoupled to give two second order ordinary differential equations for each wavefunction
\begin{align}\label{eqn:eomsmaj2nd}
 \(\partial^2_{t} +  \omega^2_\lambda(t)- i \partial_t \tilde{k}_{\lambda}(t)\)X^{\lambda}_{i k}(t)  = 0, \quad
\(\partial_{t}^2+\omega^2_\lambda(t)  + i\partial_{t}\tilde{k}_{\lambda}(t) \)Y^{\lambda *}_{i k}(t)  = 0,
\end{align}
where
\begin{align}\label{eqn:fermionfreq}
\omega^2_\lambda(t) = &  \tilde{k}_{\lambda}(t)^2 + m_i^2,\quad \tilde{k}_{\lambda}(t) =  \(\frac{k}{a}\lambda +\frac{C}{f}\dot\phi \).
\end{align}

We are interested in particle creation, and so we compute the Hamiltonian density 
\begin{align}
\mathcal{H}_{\rm f} = \pi^\beta_{\chi_i} \dot\chi_{i \beta} - \mathcal{L}_{\rm f} =  -  a^{-1} i\chi^\dagger_{i \dot\alpha} \bar{\sigma}^{j \dot\alpha \beta}\partial_{j} \chi_{i \beta} &  + \frac{m_{i}}{2}(\chi_i ^{\alpha}\chi_{i\alpha} + \chi^{\dagger}_{ i \dot\alpha}\chi_i^\dagger {}^{\dot\alpha} )  -\frac{C}{f}\dot\phi \chi^\dagger_{i \dot\alpha} \bar{\sigma}^{0 \dot\alpha \beta} \chi_{i \beta}.
\end{align}
After Fourier transforming, inserting the mode expansions, and some straightforward algebra, the Hamiltonian is given by
\begin{align}\label{eqn:majmodehamiltonian}
H_{\rm f}
 = \sum_\lambda \int  \frac{{\rm d}^3 k}{(2\pi)^3}\, \Bigg\{ & E_i^\lambda(k) a^\dagger{}^\lambda _{\bf k} a_{\bf k}^{\lambda} +F_i^\lambda(k) \lambda \iota^{\lambda}_{\hat{k}}  a^\dagger{}^\lambda _{-\bf k} a^\dagger{}^\lambda _{\bf k} 
 +F^{\lambda*}_i(k)  \lambda \iota^{\lambda*}_{\hat{k}} a_{\bf k}^\lambda a_{-\bf k}^\lambda 
- E^{\rm vac}_\lambda(k)(2\pi)^{3}\delta^{3}(0)\Bigg\},
\end{align}
where
\begin{align}\nn\label{eqn:majenergyeqs}
E_i^\lambda(k) = &\tilde{k}_{\lambda}(Y^{\lambda *}_{i k}(t) Y^{\lambda}_{i k}(t) - X^{\lambda *}_{i k} (t)X^{\lambda}_{i k} (t)) +m_i  (X^{\lambda *}_{i k} (t) Y^{\lambda*}_{i k}(t)+Y^{\lambda }_{i k}(t)  X^{\lambda}_{i k} (t)),
\\\nn
 F_i^\lambda(k) = & \tilde{k}_{\lambda}  X^{\lambda *}_{i k} (t)Y^{\lambda }_{i k}(t) - \frac{m_i}{2}(Y^{\lambda}_{i k}(t)  
Y^{\lambda}_{i k}(t)  -X^{\lambda *}_{i k} (t) X^{\lambda*}_{i k} (t)   ),  \\
E^{\rm vac}_\lambda(k) = & \tilde{k}_{\lambda}Y^{\lambda *}_{i k}(t) Y^{\lambda}_{i k}(t) + \frac{m_i}{2}(X^{\lambda }_{i k} (t) Y^{\lambda}_{i k}(t) + Y^{\lambda *}_{i k}(t)X^{\lambda*}_{i k} (t))  .
 \end{align}
Note that the presence of the phase $\iota^{\lambda}_{\hat{k}}$ prevents the off-diagonal terms in the Hamiltonian in Eqn.\ \eqref{eqn:majmodehamiltonian} from vanishing. 
The diagonal and off-diagonal parts of the Hamiltonian are related by
\begin{align}
(E_i^\lambda)^2 + 4 |F_i^\lambda|^2 = \left ( \tilde k_\lambda ^2 + m_i^2 \right ) \left ( X_{ik}^\lambda X_{ik}^{\lambda*}  +   Y_{ik}^{\lambda}  Y_{ik}^{\lambda*}   \right )^2 = {\omega^2_\lambda \over 4}.
\end{align}

\subsection{Degenerate fermions: Dirac fermions}

In the case where the mass matrix has degenerate eigenvalues there is a global internal $O(2)$ flavor symmetry $\f_i \to \mathcal{O}_i{}^{j}\f_j$, where $\mathcal{O}$ is an orthogonal matrix, provided we assume that $C_{i} = C$, that is, that the axion couples universally to the fermions.  There is a conserved hermitian Noether current associated with this symmetry
\begin{align}\label{eqn:so2noether}
J^\mu = & i \(\f^{\dagger}_{1 \dot\alpha}\bar{\sigma}^{\mu \dot\alpha \beta}\f_{2\beta}-\f^{\dagger}_{2 \dot\alpha}\bar{\sigma}^{\mu \dot\alpha \beta}\f_{1\beta}\).
\end{align}
However, the flavor basis we have been using is off-diagonal with respect to this current and thus the flavor states are not eigenstates of the charge operator $Q = \int {\rm d}^3x J^0$.  It is more useful to work in a basis of states of definite charge.   We can achieve this by diagonalizing the Noether current. Redefining our fields
\begin{align}
\varphi \equiv & \frac{1}{\sqrt{2}}(\f_1 + i \f_2), \quad \eta \equiv  \frac{1}{\sqrt{2}}(\f_1 - i \f_2),
\end{align}
diagonalizes the current in Eqn.\ \eqref{eqn:so2noether}. In this basis, the action reads
\begin{align}\nn
S_{\rm f} =  \int {\rm d}^3 x {\rm d}t \bigg[i \varphi^{\dagger}_{\dot\alpha} \bar{\sigma}^{\mu \dot\alpha \beta}\partial_{\mu} \varphi_{\beta} +i \eta^{\dagger}_{\dot{\alpha}} \bar{\sigma}^{\mu \dot\alpha \beta}\partial_{\mu} \eta_\beta & - m(\varphi^{\alpha} \eta_{\alpha}+ \varphi^{\dagger}_{\dot\alpha}\eta^{\dagger \dot\alpha})\\ &  + \frac{C}{f}\partial_{\mu}\phi( \varphi^{\dagger}_{\dot\alpha} \bar{\sigma}^{\mu \dot\alpha \beta} \varphi_{\beta} + \eta^{\dagger}_{\dot{\alpha}} \bar{\sigma}^{\mu \dot\alpha \beta} \eta_\beta)  \bigg]
\end{align}
and the $SO(2)$ flavor symmetry (the part continuously connected to the identity) is realized in this basis as the $U(1)$ symmetry $\varphi \to e^{i\theta}\varphi$ and $\eta \to e^{- i \theta}\eta$, where $\theta$ is the angle that defines the $SO(2)$ rotation matrix \cite{Dreiner:2008tw}. 

The spinors $\varphi$ and $\eta$ satisfy the free-field Dirac equations
\begin{align}
i\bar{\sigma}^{\mu \dot\alpha \beta}\partial_{\mu} \varphi_{\beta}+ \frac{C}{f}\partial_{\mu}\phi \bar{\sigma}^{\mu \dot\alpha \beta} \varphi_{\beta}  - m\eta^{\dagger \dot\alpha} = & 0,\\ 
i\bar{\sigma}^{\mu \dot\alpha \beta}\partial_{\mu} \eta_{\beta}+ \frac{C}{f}\partial_{\mu}\phi \bar{\sigma}^{\mu \dot\alpha \beta} \eta_{\beta}  - m\varphi^{\dagger \dot\alpha} = & 0.
\end{align}
Together $\varphi$ and $\eta^{\dagger}$ compose a Dirac fermion,  which we can expand into plane waves as
\begin{align}
\varphi_\alpha({\bf x}, t) = & \sum_{\lambda} \int \frac{{\rm d}^3 k}{(2\pi)^3}\[x^\lambda_{\alpha}({\bf k}, t) a_{\bf k}^\lambda e^{i {\bf k}\cdot x} + y^\lambda_{\alpha}({\bf k}, t) b^{\lambda\dagger}_{\bf k} e^{-i {\bf k}\cdot x}\],\\
\eta_\alpha({\bf x}, t) = & \sum_{\lambda} \int \frac{{\rm d}^3 k}{(2\pi)^3}\[x^\lambda_{\alpha}({\bf k}, t) b_{\bf k}^\lambda e^{i {\bf k}\cdot x} + y^\lambda_{\alpha}({\bf k}, t) a^{\lambda\dagger}_{\bf k} e^{-i {\bf k}\cdot x}\],
\end{align}
where $a^\dagger, b^\dagger, a$, and $b$ are creation and annihilation operators that satisfy the anticommutation relations
\begin{align}
\{a_{\bf k}^\lambda ,  a_{\bf k'}^{\dagger \lambda'}\} = & \{b_{\bf k}^\lambda ,  b_{\bf k'}^{\dagger \lambda'}\} = \delta_{\lambda \lambda'}\delta^{3}({\bf k} - {\bf k}'),
\end{align}
with all others vanishing. It follows that
\begin{align}
\varphi^\dagger_{\dot\alpha}({\bf x}, t) = & \sum_{\lambda} \int \frac{{\rm d}^3 k}{(2\pi)^3}\[x^{\lambda\dagger}_{\dot\alpha}({\bf k}, t) a^{\lambda\dagger}_{\bf k}e^{-i {\bf k}\cdot x} + y^{\lambda\dagger}_{\dot\alpha}({\bf k}, t) b^{\lambda}_{\bf k} e^{i {\bf k}\cdot x}\],\\
\eta^\dagger_{\dot\alpha}({\bf x}, t) = & \sum_{\lambda} \int \frac{{\rm d}^3 k}{(2\pi)^3}\[x^{\lambda\dagger}_{{\dot\alpha}}({\bf k}, t) b^{\lambda\dagger}_{\bf k} e^{-i {\bf k}\cdot x} + y^\dagger_{{\dot\alpha}}({\bf k}, t) a^{\lambda}_{\bf k} e^{i {\bf k}\cdot x}\].
\end{align}
Substituting these mode expansions into the equations of motion, we find the same equations of motion as in the non-degenerate Majorana case 
\begin{align}
 i \(\bar{\sigma}^{0 \dot{\alpha}\beta}\partial_{0}+ i \frac{k_j}{a} \bar{\sigma}^{j \dot{\alpha}\beta}\)x^\lambda _{ \beta}({\bf k}, t)+\frac{C}{f}\dot\phi \bar{\sigma}^{0 \dot\alpha \beta} x^\lambda_{\beta} ({\bf k}, t) = & m y^{\lambda \dagger\dot\alpha}({\bf k}, t), \\
 i \({\sigma}^{0}_{ \delta \dot\gamma}\partial_{0} + i \frac{k_j}{a} {\sigma}^{j}_{ \delta\dot\gamma}\)y^{\lambda  \dagger \dot\gamma}({\bf k}, t)-\frac{C}{f}\dot\phi {\sigma}^{0}_{ \delta\dot\gamma} y^{\lambda  \dagger \dot\gamma}({\bf k}, t)= & m x^\lambda _{\delta}({\bf k}, t).
\end{align}
We proceed in the same fashion as the non-degenerate mass case, and work in a basis of helicity eigenspinors, as above
\begin{align}
x_{\delta}^{\lambda}({\bf k}, t) = & X^{\lambda}_{k} (t) \xi_\lambda({\bf k}), \quad 
y^{\lambda \dagger \dot\alpha }({\bf k}, t) =  Y^{\lambda *}_{ k}(t) \xi_\lambda({\bf k}).
\end{align}
Then, the equations of motion for the wavefunctions of the Dirac fields are identical to Eqn.\ \eqref{eqn:eomsmaj1st} derived earlier for each flavor in the non-degenerate system,
\begin{align}
i \(\partial_{t}- i  \frac{k}{a}\lambda  \)X^{\lambda}_{ k} (t) +\frac{C}{f}\dot\phi X^{\lambda}_{ k} (t) = & m Y^{\lambda *}_{ k}(t), \\
 i \(\partial_{t} + i  \frac{k}{a} \lambda \)Y^{\lambda*}_{k}(t)-\frac{C}{f}\dot\phi Y^{\lambda *}_{k}(t) = & m X^{\lambda}_{k} (t).
\end{align}
However, note that for two fermions there is only a single pair of equations, unlike the Majorana case where there is a pair of equations for each flavor. This is ultimately because Dirac particles are related to their antiparticles by charge conjugation; the $U(1)$ symmetry relates the fields to each other.  We can construct the Hamiltonian for this theory analogously to before
\begin{align}\label{eqn:diracmodehamiltonian}
H_{\rm f} =   \sum_{\lambda}  \int  \frac{{\rm d}^3 k}{(2\pi)^3}\Big\{ E_{\lambda}(k, t)  (a^\dagger{}^\lambda _{\bf k}  a_{\bf k}^{\lambda}+b^\dagger{}^\lambda _{\bf k}  b_{\bf k}^{\lambda}) & -  E^{\rm vac}_{\lambda}(k, t) (2\pi)^3 \delta^{3}(0)\\ \nn& 
 +F_{\lambda}(k, t)\lambda \iota^{\lambda}_{\hat{k}}   b^\dagger{}^{\lambda} _{-\bf k}a^\dagger{}^\lambda _{\bf k}
 + F^*_{\lambda}(k, t)  \lambda \iota^{\lambda *}_{\hat{k}} a_{\bf k}^{\lambda}b_{-\bf k}^\lambda
\Big\},
\end{align}
where we have defined
\begin{align}\nn \label{eqn:diracenergyeqs}
E_{\lambda}(k) = & \tilde{k}_{\lambda}  (  Y^{\lambda *}_{k}(t)   Y^{\lambda}_{ k}(t)-X^{\lambda *}_{k} (t)    X^{\lambda}_{ k} (t))+m (X^{\lambda *}_{ k} (t)  Y^{\lambda*}_{ k}(t)+X^{\lambda }_{ k} (t)  Y^{\lambda}_{ k}(t)),\\\nn
F_{\lambda}(k) = &  2 \tilde{k}_{\lambda}   X^{\lambda *}_{k} (t) Y^{\lambda}_{ k}(t) - m (Y^{\lambda }_{ k}(t)  
 Y^{\lambda}_{k}(t)-X^{\lambda *}_{ k} (t) X^{\lambda*}_{ k} (t)), \\
E^{\rm vac}_{\lambda}(k) = & 2\tilde{k}_{\lambda}  Y^{\lambda *}_{ k}(t)   Y^{\lambda}_{ k}(t) + X^{\lambda }_{ k} (t)  Y^{\lambda}_{ k}(t)+Y^{\lambda *}_{ k}(t) X^{\lambda *}_{ k} (t)),
\end{align}
which is simply the sum over the flavors of the Hamiltonian at Eqn.\ \eqref{eqn:majmodehamiltonian} where $m_i = m$.
The relations between these functions for Majorana and Dirac Hamiltonians are
\begin{align}
E_{\lambda}^{\rm Dirac}(k)  =E_{\lambda}^{\rm Majorana}(k), \\
F_{\lambda}^{\rm Dirac}(k)  = 2 F_{\lambda}^{\rm Majorana}(k) ,
\end{align}
and the corresponding consistency relation in the Dirac case becomes
\begin{align}
(E_\lambda)^2 +  |F_\lambda|^2 = \left ( \tilde k_\lambda ^2 + m_i^2 \right ) \left ( X_{ik}^\lambda X_{ik}^{\lambda*}  +   Y_{ik}^{\lambda}  Y_{ik}^{\lambda*}   \right )^2 ={\omega^2 _\lambda \over 4}.
\end{align}

\subsection{Vacuum states}

The Hamiltonian in both Eqn.\ \eqref{eqn:majmodehamiltonian}  and Eqn.\ \eqref{eqn:diracmodehamiltonian}  can be diagonalized at any given time $t_0$ by choosing the wavefunctions $X^\lambda_k$ and $Y^\lambda_k$ so that the function $F^\lambda_k$ vanishes. A straightforward calculation shows that $F^\lambda_k = 0$ whenever
\begin{align}
X^\lambda_k \propto &\(1\mp \frac{\tilde k_\lambda(t)}{\omega_\lambda(t)}\)^{1/2} e^{i\phi},  \quad
Y^\lambda_k \propto \(1\pm \frac{\tilde k_\lambda(t)}{\omega_\lambda(t)}\)^{1/2} e^{-i\phi},
\end{align}
where $\phi$ is an arbitrary phase. Here, we are interested in modes that start in the Bunch-Davies vacuum state during inflation. In order to determine the initial condition, we look for Wentzel-Kramers-Brillouin (WKB) solutions to the equations of motion. Assuming that 
$
|\dot\omega/\omega^2| \ll 1,
$
it is straightforward to derive the lowest order WKB solution to Eqs.\ \eqref{eqn:eomsmaj1st} - \eqref{eqn:eomsmaj2nd}
\begin{align}\nn\label{eqn:WKBbasis}
X_{i k}^\lambda(t) = A_{i\lambda} \sqrt{1+\frac{ \tilde{k}_\lambda}{\omega_\lambda}}e^{i \int\omega_\lambda dt} + B_{i\lambda} \sqrt{1-\frac{ \tilde{k}_\lambda}{\omega_\lambda}}e^{- i \int\omega_\lambda dt},\\
Y_{i k}^{\lambda*}(t) = - A_{i\lambda} \sqrt{1-\frac{ \tilde{k}_\lambda}{\omega_\lambda}}e^{i \int\omega_\lambda dt} + B_{i\lambda} \sqrt{1+\frac{ \tilde{k}_\lambda}{\omega_\lambda}}e^{- i \int\omega_\lambda dt},
\end{align}
where as above
\begin{align}
\omega^2_\lambda(t) = &  \tilde{k}_{\lambda}(t)^2 + m_i^2,\quad \tilde{k}_{\lambda}(t) =  \(\frac{k}{a}\lambda +\frac{C}{f}\dot\phi \).
\end{align}
Substituting these solutions into Eqn.\ \eqref{eqn:majenergyeqs} we obtain
\begin{align}\nn\label{eqn:WKBenergys1}
E^\lambda_i(k) = & 2\omega_\lambda (  |B_{i\lambda}|^2 - |A_{i\lambda}|^2),\quad 
 F^\lambda_i(k) =   2\omega A^*_{i\lambda} B^*_{i\lambda},\\ %
E^{\rm vac}_\lambda(k) = &  \tilde{k}_\lambda(|A_{i\lambda}|^2 + |B_{i\lambda}|^2  ) + \omega_\lambda (  |B_{i\lambda}|^2 - |A_{i\lambda}|^2),
\end{align} 
for the Majorana case, while substituting them into Eqn.\ \eqref{eqn:diracenergyeqs} gives
\begin{align}\nn\label{eqn:WKBenergys2}
E^\lambda_i(k) = & 2\omega_\lambda (  |B_{i\lambda}|^2 - |A_{i\lambda}|^2),\quad
 F^\lambda_i(k) =   4\omega A^*_{i\lambda} B^*_{i\lambda},\\
E^{\rm vac}_\lambda(k) = & 2 \tilde{k}_\lambda(|A_{i\lambda}|^2 + |B_{i\lambda}|^2  ) + 2\omega_\lambda (  |B_{i\lambda}|^2 - |A_{i\lambda}|^2),
\end{align} 
for the Dirac case. On the other hand, canonical quantization requires that the coefficients satisfy
\begin{align}
\sum_\lambda\[ |A_{i\lambda}|^2 + |B_{i\lambda}|^2   \] = {1\over 2}.
\end{align}
Thus, in order for the fields to be canonically normalized and in the lowest energy state (the state with the lowest vacuum energy density), we are required to take 
\begin{align}\label{eqn:initialconditions}
A_{i \lambda} = {1\over 2}, \quad B_{i\lambda} = 0. 
\end{align}
Notice that in the absence of the coupling to the axion (when $\tilde{k}_{\lambda} = k\lambda/a$), the first term in the vacuum energy in Eqn.\ \eqref{eqn:WKBenergys1} - \eqref{eqn:WKBenergys2} vanishes when summed over spins. When the coupling to the axion is turned on, this term no longer vanishes, and leads to a splitting of the vacuum levels for the $\lambda = +1$ and $\lambda = -1$ helicity states.

The choice of initial conditions at Eqn.\ \eqref{eqn:initialconditions} diagonalizes the Hamiltonian at some initial time, but in general the equations of motion drive $F^{\lambda}_i(k)$ away from zero at times $t>t_0$. At these later times, the Hamiltonian can be re-diagonalized after performing a (time-dependent) Bogoliubov transformation on the creation and annihilation operators.

We begin by constructing the Bogoliubov transformation for the Majorana case, by first writing the Hamiltonian in the following form
\begin{align}
H_{\rm f} =  \sum_\lambda \int {{\rm d}^3k\over (2\pi)^3 }  \left \{
\begin{pmatrix}a_k ^\dagger & a_{-k} \end{pmatrix}
  \begin{pmatrix} E^\lambda/2 & -F_\lambda(k) \lambda \iota_k^\lambda  \\ -F_{\lambda }^*(k) \lambda \iota^{\lambda *}_k   & -E^\lambda/2  \end{pmatrix} 
  \begin{pmatrix} a_k \\ a_{-k}^\dagger \end{pmatrix} - E_{\lambda}^{\rm vac} (2\pi)^3 \delta^3(0) \right \}.
\end{align}
This Hamiltonian can be diagonalized by choosing
\begin{align}
\tilde{a}_{\bf k}^{\lambda} = & \alpha_k(t) a^\lambda_{\bf k} + \beta_{k}(t) \iota_{-k} a^{\dagger \lambda}_{-\bf k},
\end{align}
where $\alpha_k(t)$ and $\beta_k(t)$ are complex functions of time that depend on the magnitude of the wavenumber. Under this Bogoliubov transformation, the creation-annihilation operator pairs become 
\begin{align}
\nn
& \tilde a_k^\dagger \tilde a_k  = |\alpha|^2 a_k^\dagger a_k + |\beta|^2 (\iota_{-k} \iota^*_{-k}) a_{-k} a_{-k}^\dagger + \alpha^* \beta \iota_{-k} a_k^\dagger a_{-k}^\dagger + \beta^* \alpha  \iota^* _{-k} a_{-k} a_k,
 \\
 \nn
& \tilde a_{-k}^\dagger \tilde a_k ^\dagger = (\alpha^*)^2 a_{-k}^\dagger a_k ^\dagger + (\beta^*)^2 (\iota_{-k}^* \iota_{k}^* )a_k a_{-k} + \alpha^* \beta^* \iota^*_{-k} a_{-k}^\dagger a_{-k}^\dagger a_{-k} + \beta^* \alpha^* \iota^*_k a_k a_k^\dagger,
 \\
& \tilde a_k \tilde a_{-k} =  \alpha^2 a_{k} a_{-k}  + \beta^2 (\iota_k \iota_{-k}) a_{-k}^\dagger a_{k}^\dagger + \alpha\beta \iota_k a_k a_k^\dagger+ \beta \alpha \iota_{-k}  a_{-k}^\dagger a_{-k}.
 \end{align}
The off-diagonal part proportional to $a_{k}a_{-k}$ is
\begin{align}\label{eqn:Hdiageqn}
-E \beta^* \alpha \iota_{-k}^* + F (\beta^*)^2 (\iota^*_{-k}\iota^*_k) + F^* \alpha^2  \, ,
\end{align}
where we supressed helicity and flavor indices. Diagonalizing the Hamiltonian amounts to setting the off-diagonal terms equal to zero. Solving for $\alpha$ is now straightforward, setting Eqn.\ \eqref{eqn:Hdiageqn} to zero yields
\begin{align}
\alpha  ={\beta^*  \iota_{-k}^* \over 2F^*} \left ( E   + {\omega\over 2} \right )  ,
\end{align}
where we used $\iota_{-k} = - \iota_k$, and took $E^2 + 4|F|^2 = {\omega^2/ 4}$.
Using the relation 
$
|\alpha|^2 + |\beta|^2 =1
$
 leads to
\begin{align}
 |\beta|^2 = {8|F|^2 \over \omega(2E+\omega) }= 4|B|^2,
\end{align}
where we used the fact that $|A_{i\lambda}|^2 + |B_{i\lambda}|^2 = 1/4$ when not summed over spins.

We can proceed in a similar fashion for the Dirac case, where the Hamiltonian can be written as
\begin{align}
H_{\rm f} = \sum_\lambda \int {{\rm d}^3k \over (2\pi)^3}  \left \{ \begin{pmatrix}a_k ^\dagger & b_{-k} \end{pmatrix} H_{2 \times 2 } \begin{pmatrix} a_k \\ b_{-k}^\dagger \end{pmatrix}  + \begin{pmatrix}b_k ^\dagger & a_{-k} \end{pmatrix} H_{2 \times 2 } \begin{pmatrix} b_k \\ a_{-k}^\dagger \end{pmatrix} 
 - E_{\rm vac}^\lambda (2\pi)^3\delta^{3}(0) \right \},
\end{align}
where
\begin{align}
H_{2 \times 2 } = 
 \begin{pmatrix} E^\lambda/2 & -F_\lambda(k) \lambda \iota_k^\lambda  \\ -F_{\lambda }^*(k) \lambda \iota^{\lambda *}_k   & -E^\lambda/2  \end{pmatrix} .
\end{align}

Note that the Dirac Hamiltonian is comprised of two identical copies of the Majorana Hamiltonian, as expected. The form of the Bogoliubov transformation can be read off as
\begin{align}
\tilde{a}_{\bf k}^{\lambda} = & \alpha_k(t) a^\lambda_{\bf k} + \beta_{k}(t)b^{\dagger \lambda}_{-\bf k},\\
\tilde{b}_{\bf - k}^{^\dagger\lambda} = & -\beta^*_k(t) a_{\bf k}^\lambda + \alpha_{k}^*(t)b^{\dagger^\lambda}_{-\bf k}\, ,
\end{align}
as, for example, in Ref.\ \cite{Peloso:2000hy}. The complex phases $\iota_k$ are absorbed into the parameter $\beta$ in this case, since -- in contrast to the Majorana case -- we do not need to invert the wavenumber of the operators as defined in the Bogoliubov transformation, so the antisymmetric nature of $\iota_k$ for the relabelling ${\bf k} \to -{\bf k}$ is irrelevant.

Performing the Bogoliubov transformations on each part of the Dirac Hamiltonian, we obtain (suppressing flavor and helicity indices for clarity) 
\begin{align}
\nn
&\tilde a^\dagger_k \tilde a_k = |\alpha|^2 a_k^\dagger a_k + |\beta|^2 b_{-k} b_{-k}^\dagger + \alpha^* \beta a_k^\dagger b_{-k}^\dagger + \alpha \beta^* b_{-k} a_k,
\\
\nn
&\tilde b^\dagger_{-k}  \tilde b_{-k}  =  |\beta|^2 a_{k} a_{k}^\dagger + |\alpha|^2 b_{-k}^\dagger b_{-k} - \beta^* \alpha a_k b_{-k} -\beta \alpha^* b_{-k}^\dagger a_k^\dagger,
\\
\nn
&\tilde b_{-k}\tilde a_k= -\alpha \beta a_{k}^\dagger a_k +\alpha\beta b_{-k} b_{-k}^\dagger - \beta^2 a_{k}^\dagger b_{-k}^\dagger + \alpha^2 b_{-k} a_k,
\\
&\tilde a_k^\dagger \tilde b_{-k}^\dagger = -\alpha^* \beta^* a_k^\dagger a_{k} + \beta^* a^*b_{-k}b_{-k}^\dagger + (\alpha^*)^2 a_k^\dagger b_{-k}^\dagger -(\beta^*)^2 b_{-k}a_{k}.
\end{align}

The off-diagonal terms (proportional to $b_{-k} a_k$) in this expression are
\begin{align}
(2E \alpha \beta^*  + F a^2  - F^*(\beta^*)^2) b_{-k} a_{k}.
\end{align}
Diagonalizing the Hamiltonian requires that these terms vanish and gives
\begin{align}
 \alpha = {-(E+\omega/2) \over F} \beta^*.
\end{align}
Using the relation $|\alpha|^2+|\beta|^2=1$ we get 
\begin{align}
|\beta|^2 = 4|B|^2,
\end{align}
as in the Majorana case.

We use these time-dependent operators (the $\tilde{a}^\dagger$ and $\tilde{b}^\dagger$) to define (time-dependent) Fock spaces built from the zero (quasi)particle states
\begin{align}
\tilde{a}_{\bf k}^{\lambda}(t)|0,t\rangle = \tilde{b}_{\bf k}^{\lambda}(t)|0,t\rangle = 0.
\end{align}
As noted above, a state that starts off at time $t_0$ as the vacuum state, i.e. a state containing no particles, evolves to a state which at a generic later time contains particles. This is because the particle number at a given moment is provided by operators which instantaneously diagonalize the Hamiltonian.
This instantaneous particle number can be computed by projecting the full mode-functions onto an instantaneous WKB basis (Eqn.\ \eqref{eqn:WKBbasis}), and extracting the negative frequency parts.

For both Majorana and Dirac Fermions, the particle number can be evaluated as
\begin{align}
n_k ^\lambda = |\beta_k ^\lambda| ^2 =  4 |B_k ^\lambda|^2.
\end{align}
We express this quasi-particle number in a more useful form, by noting that the particle number	can be extracted by comparing a given solution of the Dirac equation to an instantaneous WKB solution of the form Eqn.\ \eqref{eqn:WKBbasis}. The square of the coefficient of the negative frequency mode is the particle number and can be extracted via
\begin{align}\label{eqn:quasiparticlenum}
n_{ik}^\lambda =\frac{1}{\omega_\lambda(\tilde{k}_\lambda + \omega_\lambda)}\[|\dot{X}^\lambda_{ik}|^2 + \omega_{\lambda}^2|{X}^\lambda_{ik}| - 2 \omega_\lambda \Im (X^\lambda_{ik}\dot{X}^{\lambda*}_{ik})\].
\end{align}
Note, however, that this particle number is only well defined where the WKB approximation is valid. That is, during times where $|\dot\omega/\omega_\lambda^2| \ll 1$.

\section{Inflationary solutions}\label{sec:inflation}

During the inflationary epoch, the accelerated expansion of spacetime leads to particle production of all fields that are not conformally invariant \cite{Parker:2009uva}. In this section we derive analytic solutions for the quasi-particle number by directly solving the Dirac equation for the fermion wavefunctions.

During slow-roll inflation we can treat the inflationary spacetime as de Sitter space to a good approximation. In this section, we work with conformal time, which is taken to be a negative increasing quantity during inflation
\begin{align}
{\rm d} \tau = \frac{{\rm d} t}{a}, \quad \tau = \int \frac{{\rm d}t}{a}    \approx -\frac{1}{aH}.
\end{align} 
We assume that the axion rolls at an approximately constant rate in cosmic time during inflation $\ddot{\phi}\ll H\dot\phi$,
and we introduce the parameter
\begin{align}\label{eqn:varthetadef}
\vartheta = - \frac{C}{f} \frac{\dot\phi}{H} \approx {\rm const.}
\end{align}
In conformal time,  the equation of motion, Eqn.\ \eqref{eqn:eomsmaj2nd}, for the spinor fields becomes
\begin{align}
 \partial_{\tau}^2\Phi^{\lambda}_{ik} + \frac{i k\lambda }{-\tau}\( 1 +2i\vartheta\)\Phi^{\pm}_{ik}+\(k^2 + \frac{1}{4 \tau^2} +\frac{1}{\tau^2}\( \frac{m_{i}^2}{ H^2} +\vartheta^2\)\)\Phi^\pm_{ik} = 0,
\end{align}
where we have treated $H \approx$ constant and $\dot{\phi}\approx$ constant, dropping time derivatives of these quantities, and we have rescaled
\begin{align}
\Phi_{ik}^\lambda = & \frac{X^\lambda_{ik}}{\sqrt{a}}.
\end{align}
Redefining the time coordinate to $u = 2 i k\tau $ puts the equation into standard form of the Whittaker equation
\begin{align}
\partial_u^2 \Phi_{i k}^{\lambda}{} - \frac{\lambda}{u} \( \frac{1}{2} +i \vartheta\)\Phi_{i k}^{\lambda}+\(-\frac{1}{4} + \frac{1}{4 u^2} +\frac{1}{u^2}\( \frac{m_{i}^2}{ H^2} +\vartheta^2\)\)\Phi_{i k}^\lambda = 0,
\end{align}
with solution
\begin{align}\label{eqn:gensol}
\Phi_{ik}^{\lambda} = A^\lambda_{ik} W_{\kappa, \mu}(2 i k\tau) +  B^\lambda_{ik} W_{-\kappa, \mu}(-2 i k\tau),
\end{align}
where
\begin{align}
\kappa = & -\lambda\( \frac{1}{2} +i\vartheta\), \quad
\mu^2 =  -\( \frac{m_{i}^2}{ H^2} + \vartheta^2\).
\end{align}
We make use of the fact that in the limit $z \to \infty$, the Whittaker functions have limiting forms \cite{NISTHB}
\begin{align}
W_{\kappa, \mu}(2 i z) \sim e^{-i z} (2 iz)^\kappa,
\end{align}
while in the same limit, the WKB solutions to the Dirac equation, Eqs.\ \eqref{eqn:WKBbasis}, are
\begin{align}
\lim_{k\tau \to -\infty}\sqrt{1 - \frac{\tilde{k}_+}{\omega_+}}e^{ - i \int \omega_+ dt} =& \frac{m_i a}{\sqrt{2} k}e^{i\theta_1}e^{-\frac{\pi}{2}\vartheta}(2 i k\tau)^{-i \vartheta}e^{- i k\tau},\\
\lim_{k\tau \to -\infty}\sqrt{1 - \frac{\tilde{k}_-}{\omega_-}}e^{ - i \int \omega_- dt} =& \sqrt{2}e^{i\theta_2}e^{\frac{\pi}{2}\vartheta}(2 i k\tau)^{i \vartheta}e^{- i k\tau},
\end{align}
where $\theta_1$ and $\theta_2$ are irrelevant phases. Thus, choosing the Bunch-Davies vacuum implies we should set $B^\lambda_{ik} = 0$ in Eqn.\ \eqref{eqn:gensol}. 

Next, we find $Y^{\lambda*}_{ik}$ in terms of $X^\lambda_{ik}$ and thus  $\Phi^\lambda_{ik}$ by using the Dirac equation, Eqn.\ \eqref{eqn:eomsmaj1st}. In terms of  $\Psi^\lambda_{ik} = Y^{\lambda*}_{ik}/\sqrt{a}$ and $\Phi^\lambda_{ik}$ with $u = 2 i k\tau $,  Eqn.\ \eqref{eqn:eomsmaj1st} is
\begin{align}
\Psi^\lambda_{ik} = -\frac{i  u }{m_i /H}\(  \partial_u - \frac{1}{2u } - \frac{\lambda}{2} - i \frac{\vartheta}{u}\)\Phi^\lambda_{ik} =- \frac{i  u }{m_i /H}\(  \partial_u  +\lambda \(\frac{\kappa}{u}- \frac{1}{2}\)\)\Phi^\lambda_{ik}.
\end{align}
To solve this equation, we make use of the fact that the Whittaker functions satisfy the identities \cite{NISTHB}
\begin{align}\label{eqn:whittakerid1}
\(z \frac{d}{dz} z\)^n\(e^{\frac{1}{2}z}z^{-\kappa - 1}W_{\kappa, \mu}(z)\) = & \(\frac{1}{2}+\mu - \kappa\)_n\(\frac{1}{2}-\mu-\kappa\)_ne^{\frac{1}{2}z}z^{-\kappa}W_{\kappa - n, \mu}(z),\\\label{eqn:whittakerid2}
\(z \frac{d}{dz} z\)^n\(e^{-\frac{1}{2}z}z^{\kappa - 1}W_{\kappa, \mu}(z)\) = & (-1)^ne^{-\frac{1}{2}z}z^{\kappa+n-1}W_{\kappa + n, \mu}(z),
\end{align}
where $(a)_n$ is Pochhammer's symbol, which satisfies
\begin{align}
(a)_0 = 1, \quad (a)_n = a(a+1)(a+2)\ldots(a+n-1) = \frac{\Gamma(a+n)}{\Gamma(a)}.
\end{align}
For $n = 1$, Eqs.\ \eqref{eqn:whittakerid1} and \eqref{eqn:whittakerid2} are
\begin{align}\label{eqn:derivs}
z\(-\partial_z -\frac{1}{2}+\frac{\kappa}{z}\)W_{\kappa, \mu}(z) = &-\(\(\frac{1}{2}-\kappa\)^2 - \mu^2\)W_{\kappa-1, \mu}(z),\\\label{eqn:derivs2}
z\(-\partial_z +\frac{1}{2}-\frac{\kappa}{z}\)W_{\kappa, \mu}(z) = &W_{\kappa+1, \mu}(z),
\end{align}
and note that, for $\lambda = -1$, 
\begin{align}\label{eqn:mukapparelation}
\(\(\frac{1}{2}-\kappa\)^2 - \mu^2\) = \frac{m_{i}^2}{H^2}.
\end{align}
The full solutions to the Dirac equation that match onto the vacuum in the UV are then given by
\begin{eqnarray}
X^+_{i k}(k\tau) =   - \frac{i m_{i}}{H}  \frac{A_{ik}^+}{\sqrt{k\tau}}W_{-\frac{1}{2}-i\vartheta,\mu}(2 i k\tau), && \quad X^-_{i k}(k\tau) =   \frac{A_{ik}^-}{\sqrt{k\tau}}W_{\frac{1}{2} + i \vartheta, \mu}(2 i  k\tau), \\\nn
Y^{+*}_{ik}(k\tau) =\frac{A_{ik}^+}{\sqrt{k\tau}} W_{\frac{1}{2}-i\vartheta, \mu}(2 i  k\tau), \qquad\quad&& \quad Y^{-*}_{ik}(k\tau) = -\frac{i m_{i}}{H}  \frac{A_{ik}^-}{\sqrt{k\tau}} W_{-\frac{1}{2}+ i \vartheta, \mu}(2 i  k\tau).
%
\end{eqnarray}
It remains to fix the overall normalization of these modes according to (no sum on $i$)
\begin{align}\label{eqn:modenorm}
\sum_\lambda\[ X^{\lambda}_{i k} (t) X^{\lambda *}_{i k} (t) + Y^{\lambda *}_{i k}(t) Y^{\lambda}_{i k}(t)  \] = 1.
\end{align}
Making use of Eqs.\ \eqref{eqn:whittakerid1} and \eqref{eqn:whittakerid2}, the fact that the Whittaker functions satisfy the connection formula
\begin{align}
W_{\kappa, \mu}(z) = W_{\kappa, -\mu}(z),
\end{align}
and have Wronskian, 
\begin{align}
\mathcal W[W_{\kappa, \mu}(z), W_{-\kappa, \mu}(e^{\pm i \pi}z)] = e^{\mp i \pi \kappa},
\end{align}
the normalization condition at Eqn.\ \eqref{eqn:modenorm} reduces to the condition
\begin{align}
 |A_{ik}^+|e^{\frac{\pi}{2}\vartheta} = |A_{ik}^{-}| e^{-\frac{\pi}{2}\vartheta}= & \frac{1}{2}.
\end{align}
Therefore, our canonically normalized fermion wavefunction solutions are
\begin{eqnarray}\label{eqn:inflationsols}
X^+_{i k}(k\tau) =    - \frac{i m_{i}}{H}  \frac{e^{i\theta}e^{-\frac{\pi}{2}\vartheta}}{\sqrt{2k\tau}}W_{-\frac{1}{2}-i\vartheta,\mu}(2 i k\tau),&& \quad X^-_{i k}(k\tau) =   \frac{e^{i\theta}e^{\frac{\pi}{2}\vartheta}}{\sqrt{2k\tau}}W_{\frac{1}{2} + i \vartheta, \mu}(2 i  k\tau) ,\\\nn
Y^{+*}_{ik}(k\tau) = \frac{e^{i\theta'}e^{-\frac{\pi}{2}\vartheta}}{\sqrt{2k\tau}} W_{\frac{1}{2}-i\vartheta, \mu}(2 i  k\tau),\qquad\quad&& \quad Y^{-*}_{ik}(k\tau) = -\frac{i m_{i}}{H}  \frac{e^{i\theta'}e^{\frac{\pi}{2}\vartheta}}{\sqrt{k\tau}} W_{-\frac{1}{2}+ i \vartheta, \mu}(2 i  k\tau),
\end{eqnarray}
where $\theta$ and $\theta'$ are arbitrary phases.  

In the absence of the axion coupling, there are well-known solutions to the Dirac equation in de Sitter space, see for example Refs.\ \cite{Taub:1937zz, Lyth:1996yj, Collins:2004wj, Koksma:2009tc}. These results can be obtained after some algebra from Eqn.\ \eqref{eqn:inflationsols} by making use of the relation between the Whittaker and Hankel functions \cite{NISTHB}
\begin{align}
W_{0, \nu}(2 z) = \sqrt{\frac{\pi z}{2}} i ^{\nu+1}H^{(1)}_{\nu}(i z),
\end{align}
and the recursion relations for the Whittaker functions
\begin{align}
2\mu W_{\kappa, \mu}(z) - \sqrt{z}W_{\kappa+\frac{1}{2}, \mu+\frac{1}{2}}(z)+\sqrt{z}W_{\kappa+\frac{1}{2}, \mu-\frac{1}{2}}(z) = & 0,\\
\(\kappa - \mu - \frac{1}{2}\)\sqrt{z}W_{\kappa-\frac{1}{2}, \mu+\frac{1}{2}}(z) + 2\mu W_{\kappa, \mu}(z) - \(\kappa + \mu - \frac{1}{2}\)\sqrt{z}W_{\kappa-\frac{1}{2}, \mu-\frac{1}{2}}(z) = & 0.
\end{align}

\begin{figure}[t]
\centering
\includegraphics[width = 3in]{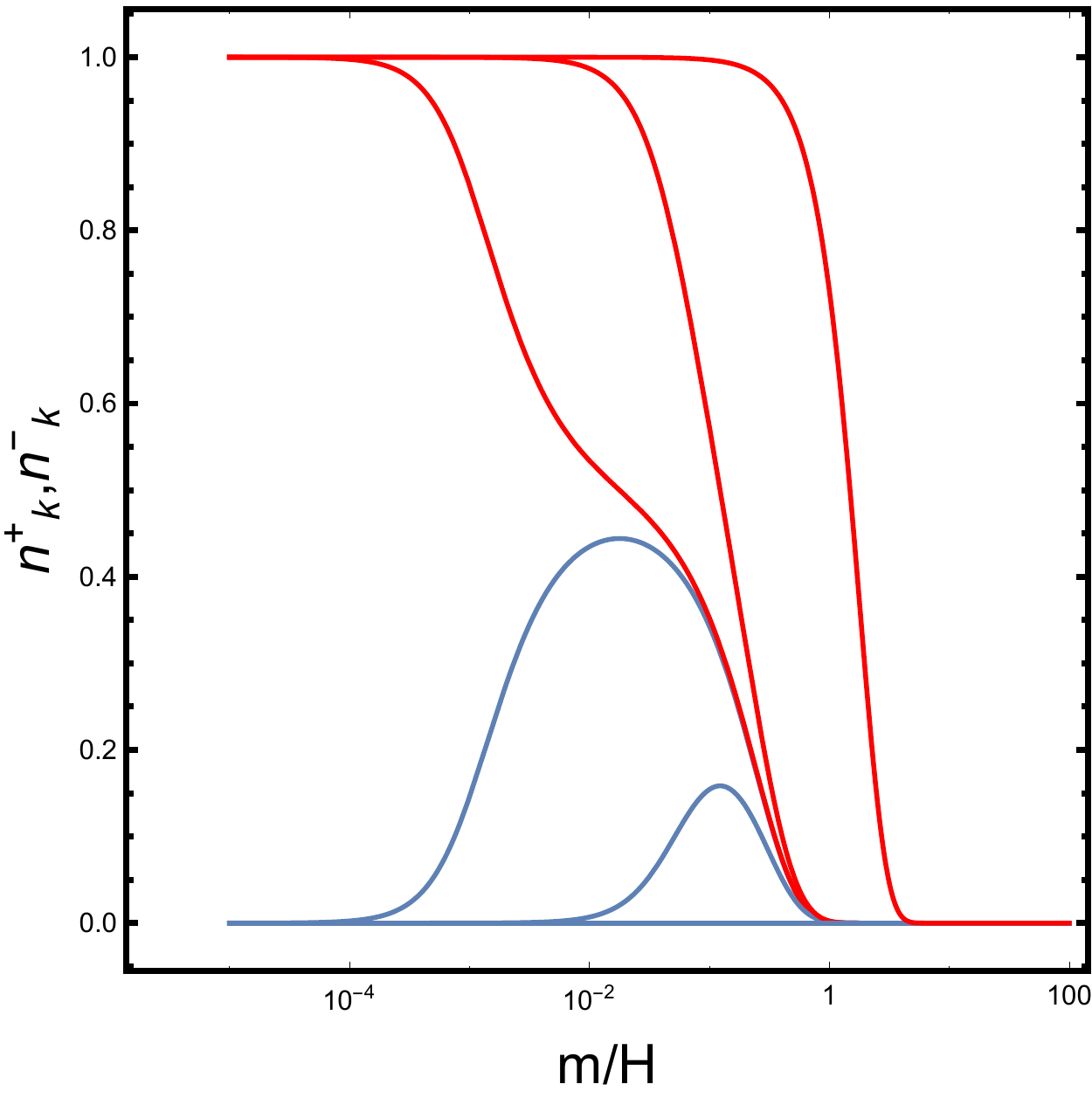}
\includegraphics[width = 3in]{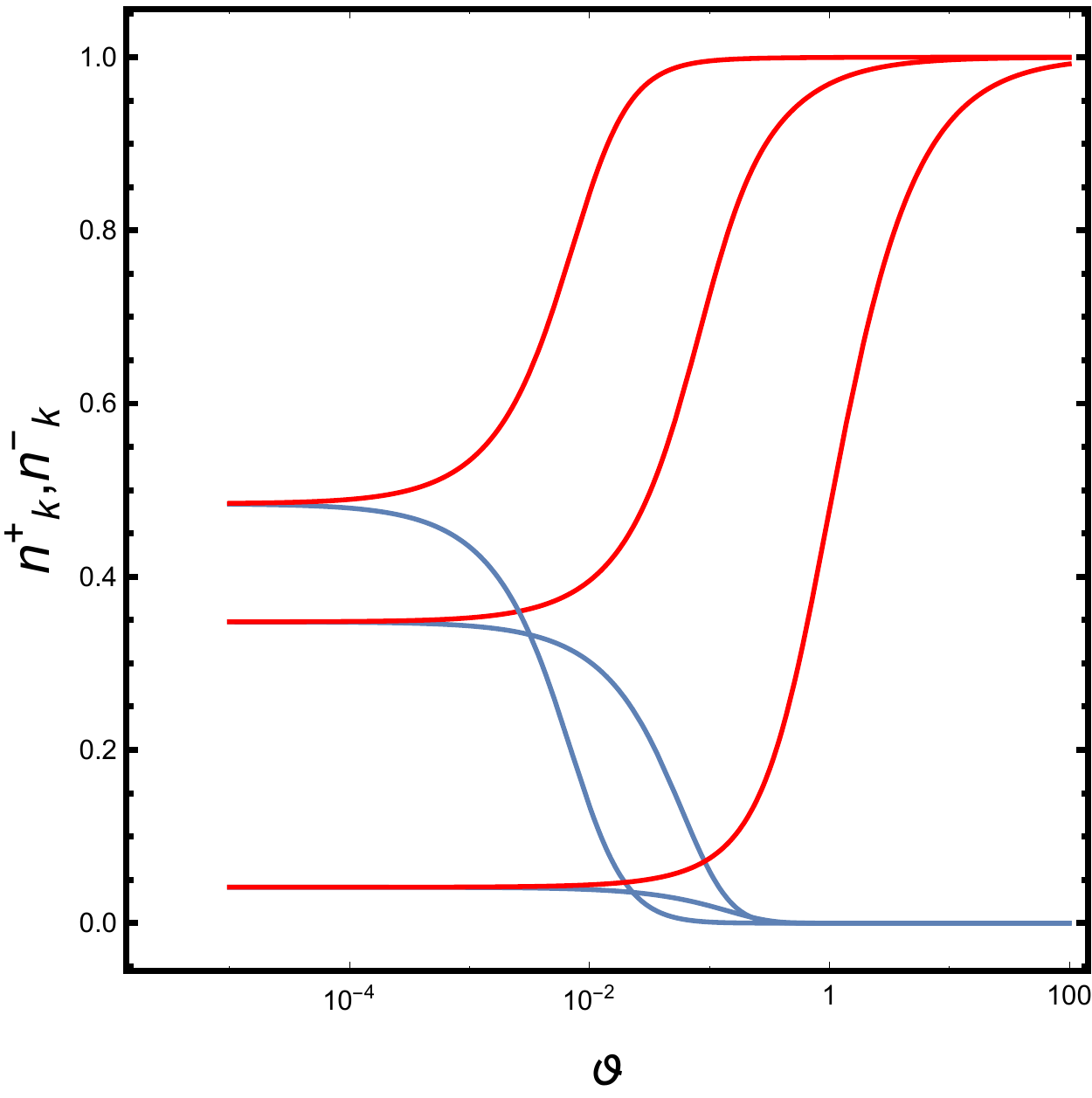}
\caption{We display the dependence of the quasi-particle number (for modes with $k < aH$) on the mass of the fermion (in units of the hubble rate) and the coupling strength to the rolling axion. In the left panel we show the particle number for fixed values of $\vartheta   = 10, 0.05, 10^{-3}$ (curves from from right to left) as the fermion mass is varied. In the right plot we show the particle number for fixed $m/H = 0.01, 0.1, 0.5$ (curves from top to bottom) as the coupling to the rolling axion is varied. In both cases, we have chosen $\vartheta > 0$ so that $n_k^{+}$ (red curves) is enhanced while $n_k^{-}$ (blue curves) is suppressed. }\label{fig:nkwithmandtheta}
\end{figure}

Using the analytic expressions in Eqn.\ \eqref{eqn:inflationsols}, we derive analytic expressions for the quasi-particle number during inflation. We are interested in modes late in inflation that have left the horizon. For modes which satisfy $k/aH \ll1$, and assuming $m_i\neq 0$, we find\footnote{When $|\vartheta|\gg m_i/H$, Eqn.\ \eqref{eqn:n_Whit} is an excellent approximation to the particle number for modes that satisfy $k/aH < |\vartheta|$.}
\begin{align}
\label{eqn:n_Whit}
n_{ik}^{\pm} = &\,e^{-\pi\(\mp\vartheta + \sqrt{\frac{m_{i}^2}{H^2}+\vartheta^2}\)}\frac{\sinh\[\pi\( \sqrt{\frac{m_{i}^2}{H^2}+\vartheta^2}\pm\vartheta\)\]}{\sinh\[2\pi\( \sqrt{\frac{m_{i}^2}{H^2}+\vartheta^2}\)\]}.
\end{align}
Eqs.\ \eqref{eqn:inflationsols} and \eqref{eqn:n_Whit} are the main result of this section.

In the absence of the coupling to the axion, production of both fermion helicities is symmetric, as expected, and highly suppressed for fermions with masses larger than the Hubble rate. For small masses, the occupation number approaches its maximum value of $1/2$ as $m_{i}/H \to 0$. However, note that for $m_i = 0$, the theory is conformally equivalent to Minkowski space, and no particle production occurs.\footnote{Eqn.\ \eqref{eqn:n_Whit} is derived using in the inflationary solutions to the Dirac equation (Eqn.\ \eqref{eqn:inflationsols}) in Eqn.\ \eqref{eqn:quasiparticlenum} and taking the limit $k\tau \to 0$. This limit does not commute with the limit $m_i \to 0$. Taking $m_i \to 0$ and then taking the limit $k\tau \to 0$, one finds the result $n_{ik}^{\pm} = 0$, as expected.} When the coupling to the axion is switched on, the particle production here is asymmetric between the helicity states. For $\vartheta > 0$ ($\vartheta < 0$), particle production of the $\lambda = +$ ($\lambda = -$) helicity state is enhanced while particle production of the $\lambda = -$ ($\lambda = +$) mode is suppressed. In Fig.\ \ref{fig:nkwithmandtheta} we display the dependence of the occupation probability on the quantity $\vartheta$, defined in Eqn.\ \eqref{eqn:varthetadef}, which encodes the effective axion-fermion coupling, and the mass of the fermion in units of the Hubble rate.

In the case where the fermions are degenerate in mass and combined to make a single Dirac fermion, we note that while the axion coupling leads to the asymmetric production of helicity states, this coupling keeps the number of particles equal to the number of anti-particles, so that the overall Noether charge is conserved. In this case, if left-handed particles are produced, so are equal numbers of right-handed anti-particles.

\section{Post-inflationary evolution and reheating in quadratic chaotic inflation}\label{sec:quadinf}

So far, we have only solved for long-wavelength modes of the fermion field in a basis of quasi-particles states. These states do not coincide with an observable basis until the Universe is in a matter- or radiation-dominated phase.\footnote{In the case where the axion coupling is absent, the particle number produced during inflation coincides with the particle number evaluated later during the matter-dominated phase that results from the axion oscillations about the minima of its potential.}  In the absence of instantaneous reheating, at the end of inflation the inflaton oscillates about the minima of its potential.  As the axion oscillates, the effective frequency of the fermion modes (see Eqn.\ \eqref{eqn:fermionfreq}) changes non-adiabatically when the effective wavenumber, $\tilde{k}_\lambda(t)$, vanishes
\begin{align}\label{eqn:ktilde}
\tilde{k}_{\lambda}(t) = \frac{k}{a}\lambda + \frac{C}{f}\dot\phi = 0.
\end{align} 
Around these times we expect particle production to occur. Note that for one of the helicity states, this non-adiabatic evolution occurs for the first time \emph{during} inflation and corresponds to the particle production considered earlier in Section \ref{sec:inflation}.  While exact analytic calculations are difficult in this regime, we can gain some insight by making use of the WKB approximation. This is the calculation we turn to next. The equations of motion for the Majorana and Dirac cases are identical, therefore we treat them uniformly, unless otherwise noted.

In order to simplify notation for the rest of this section, the comoving wavenumber $k$ and the fermion mass $m_i$ are measured in units of the axion mass $m_\phi$. Similarly,  time is measured in units of $m_\phi^{-1}$ and $f$ and $\phi$ are measured in units of $M_{\rm Pl}$. We also drop the subscript $i$ on the fermion mass, because we focus on only one flavor of fermion.

\subsection{Static universe approximation}

In order to gain intuition about the post-inflationary processes in this model, we first work in the limit where the universe does not expand.  In this static universe limit, the oscillations of the axion about the minima of its potential are harmonic, 
$\phi(t) = \phi_0 \sin( t)$,
and we can use the methods outlined by Peloso and Sorbo \cite{Peloso:2000hy} to study the evolution of the fermion particle number. Peloso and Sorbo used a Yukawa coupling of the inflaton to the fermion sector, however, the equations for the axial coupling considered here can be mapped directly onto the Yukawa case by identifying
\beq
m(t)|_{\rm Yukawa} \to \tilde k_\lambda(t)|_{\rm axion}~,~k|_{\rm Yukawa} \to  m |_{\rm axion}\, .
\eeq
Proceeding analogously to Ref.\  \cite{Peloso:2000hy} we look at the points of non-adiabatic behavior, where $\tilde k (t_*)=0$. These points occur whenever $k \lambda= -(C/f)\dot\phi$ for the wavenumber of interest. For the rest of the static universe analysis we choose $\lambda=1$ without loss of generality. We thus consider modes for which $k<{(C/ f)} \phi_0 $, where $\tilde k (t_*)=0$ is possible, and work in the limit ${C\over f} \phi_0 \gg 1$ in order for the WKB approximation to be valid.  Defining the time $t_*$ by $\tilde k(t_*)=0$, we expand the effective wavenumber near these points as
\beq
\tilde k \approx   {{\rm d} \tilde k \over {\rm d} t}   \big |_{t=t_*} (t-t_*) = -A(t-t_*),
\eeq
where $A={C\over f} \phi_0$. 
These zero-crossings happen multiple times and differ based on helicity and wavenumber, so we should be writing $t_* = t_i ^\lambda(k)$. However, we keep $t_*$, to not overly complicate our notation.
Introducing the parameters
\begin{align}
&p ={ m \over \sqrt {{C\over f} \left |  \ddot \phi_* \right |}} ={m \over  ({  {A^2 - k^2}   })^{1/4}}  , \quad 
 z =   \sqrt {{C\over f} \left |  {\ddot\phi}_* \right |} (t-t_*) = ( A^2-k^2 )^{1/4} (t-t_*),
\end{align}
the equation of motion for the fermion wavefunctions is approximated near $t_*$ as 
\begin{align}
\partial_z^2 X^{\lambda}_{i k} + (p^2 + i+ z^2)X^{\lambda}_{i k}=0 \, ,
\\
\partial_z^2 Y^{\lambda}_{i k} + (p^2 - i+ z^2)Y^{\lambda}_{i k}=0 \, .
\end{align}
If the fermion field started in the vacuum state initially ($t< t_*$), then the particle spectrum after the point where $\tilde{k} = 0$ for the first time ($t>t_*$) is given by  \cite{Peloso:2000hy}
\begin{align}
n_k ^{(1)} \approx  e^ { - \pi p_1^2}  ~,~  p_i ^2={  m^2 \over   \left |  {{\rm d}\tilde k / {\rm d}t}\right |_{t_*}   }   ,
\label{eq:nk_WKB}
\end{align}
where the superscript $(1)$ denotes the first $\tilde k$ zero-crossing, henceforth also called a production event. 
We call the combination $\pi p_i ^2$  the fermion production exponent at the $i$'th production event. This is independent of $i$ in the static universe case, but not once we take into account the effects of the expansion.
In the static universe approximation, the particle number after the first production event becomes
\begin{align}
n_k^{(1)} =  \left\{ \begin{array}{ll}\exp\left( -\pi{ m^2 \over  \sqrt{A^2-k^2} }\right ) & , k < A = \frac{C}{f}\phi_0 ,\\ 0 & , k> A = \frac{C}{f}\phi_0. \end{array}\right.
\end{align}

Fermion production is exponentially sensitive to the square of the fermion mass and the inverse of the axion-fermion coupling constant. Production is therefore significant when ${\pi m^2 / A}\lesssim 1$ and effectively shuts off when ${\pi m^2 / A}\gtrsim 3 $. A important characteristic of the present model is that each wavenumber undergoes particle production at a different time, defined by $\tilde k =0$, which for the static universe case is simply $t_* = \arccos\left (  -k/A \right )$. In the  Yukawa-coupled model studied in Ref.\ \cite{Peloso:2000hy}, particle production happens uniformily for all wavenumbers $k$ when the effective fermion mass crosses zero.

We have thus far only considered the first instance of fermion production where many modes are uniformly populated up to a maximum wavenumber set by the axion coupling. There are many oscillations of the background axion field, and we are primarily interested in the occupation numbers after a very long time. For bosons, occupation numbers grow exponentially due to Bose-enhancement and parametric resonance. However, fermions obey Fermi-Dirac statistics which limits their occupation number to be less than unity. Parametric resonance\footnote{It is true that parametric resonance is usually associated with exponential growth which can occur for Bose fields. We use the term to describe the fermionic case, acknowledging the significant differences due to Pauli blocking, in the spirit of the discussion found, for example, in \cite{Greene:2000ew}.} and Pauli blocking in the fermion case leads to chaotic filling and emptying of fermion states. In the static universe approach, the time averaged particle number count can be evaluated semi-analytically.  Following the analysis of Green and Kofman \cite{Greene:2000ew,Greene:1998nh}, we define a locally averaged particle number count 
\beq
\bar n_k(t) = {1\over T} \int_t ^{t+T} n_k(t) {\rm d} t = F_k \sin^2(\nu_k t) \, ,
\label{eq:GreenAverage1}
\eeq
where $T$ is the period of $\phi(t)$. The amplitude $F_k$ and frequency $\nu_k$ are
\beqn
F_k &=& {1\over \sin^2 (\nu_k T)} {k^2 \over 2\Omega^2_k } \left ( \Im({\cal Z}_k^{(1)} (T) \right)^2, \quad
\cos(\nu_k T) = -\Re \left ( {\cal Z}_k^{(1)} (T) \right ) \, .
\label{eq:GreenAverage2}
\eeqn
The function ${\cal Z}_k^{(1)}(t)$ is the first fundamental solution of the equation of motion, meaning that $\left . {\cal Z}_k^{(1)}\right |_{t=0}=1$ and $\left . \partial_t {\cal Z}_k^{(1)}\right |_{t=0}=0$. 

\begin{figure}[t]
\centering
\includegraphics[width=3in]{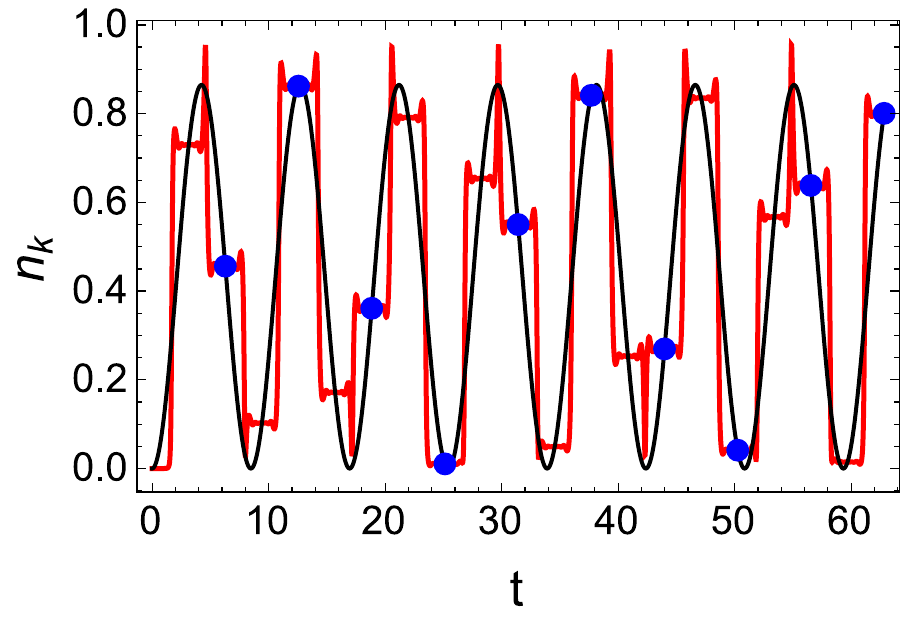} \includegraphics[width=3in]{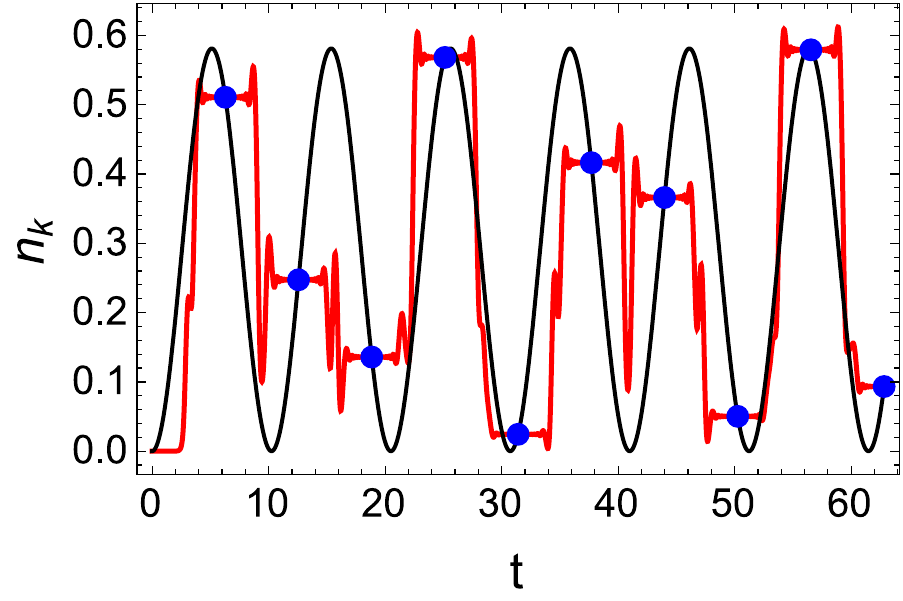}
\\
\includegraphics[width=3in]{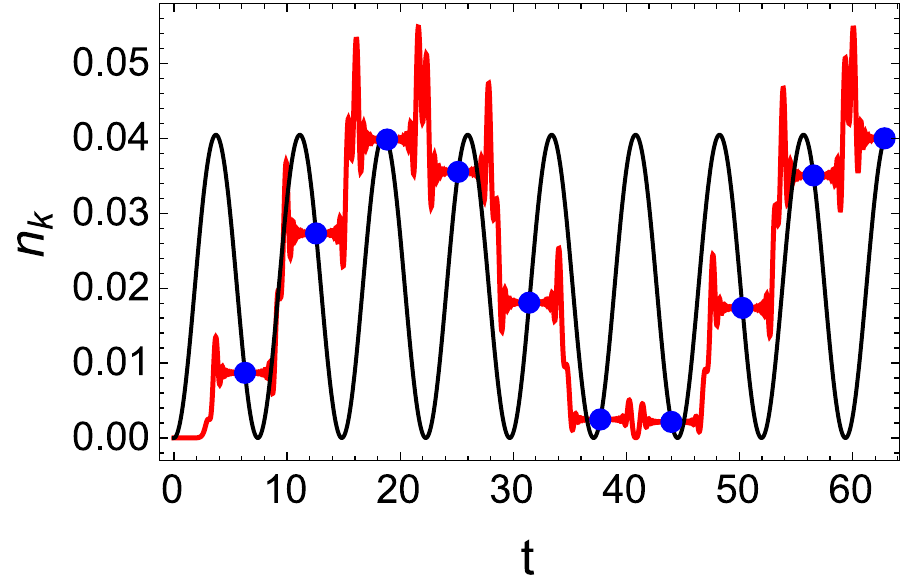} \includegraphics[width=3in]{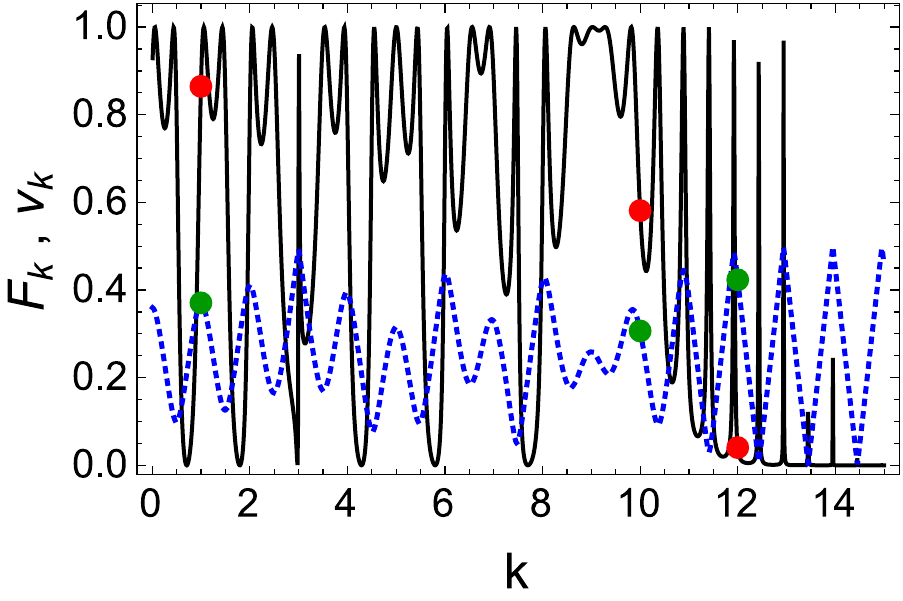}
\caption{Exact (red) and approximate (red) particle number for $m=1, {C\over f} \phi_0 = 10$ and $k=1,10,12$ (top left, top right and bottom left respectively). The blue dots correspond to $t$ being an integer multiple of $2\pi$. The values of $F_k$ (black solid) and $\nu_k$ (blue dotted) are shown in the bottom right figure. The three pairs of dots show the values of $F_k$ (red dot) and $\nu_k$ (green dot) that correspond to the three panels of this Figure.}
\label{fig:envelopebroad}
\end{figure}
\begin{figure}[t]
\centering
\includegraphics[width=3in]{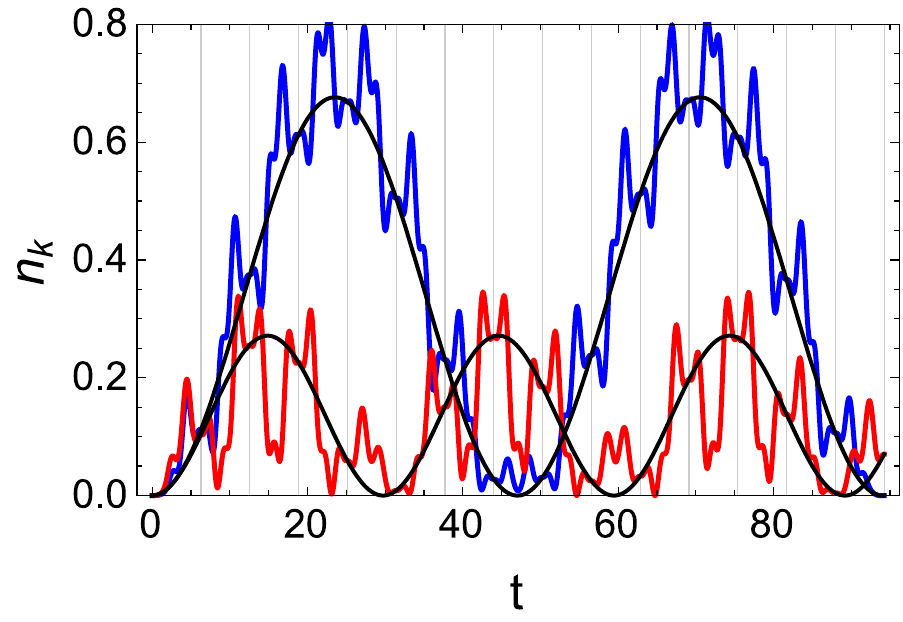} \includegraphics[width=3in]{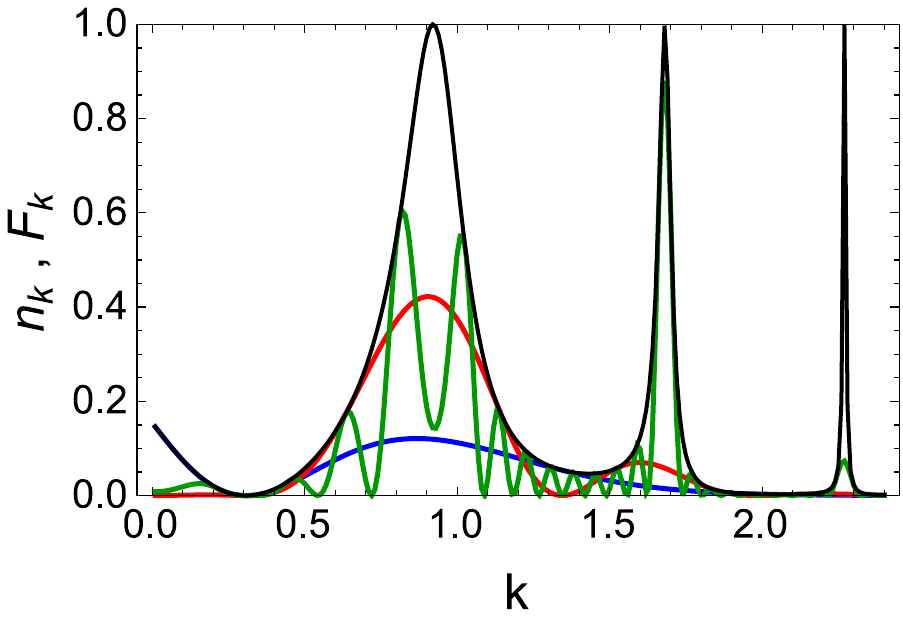}
\caption{Left panel: Evolution of the particle number for $m=1, {C\over f} \phi_0 = 1$ and $k=1$  (blue), $k=0.7$ (red). The black lines show the particle number envelope given by Eqn.\ \eqref{eq:GreenAverage1}, \eqref{eq:GreenAverage2}. The vertical lines correspond to points where $t$ is an integer multiple of $2\pi$. Right panel: Envelope function $F_k$ (black) for $m=1,{C\over f}\phi_0=1$ along with the particle spectra for $t=2\pi$ (blue), $t=4\pi$ (red) and $t=20\pi$ (green).  }
\label{fig:envelope}
\end{figure}
The WKB method used in Ref.\ \cite{Peloso:2000hy} provides a different construction for the particle number after one full inflaton oscillation. The two different analytical approximations to the fermion number produce a final result of the same form $n_k  \approx F_k \sin^2 (\nu_k t)$. However, these analytic approximations are originally constructed for different regimes. Peloso and Sorbo worked in the broad resonance regime\footnote{We define the terms broad and narrow resonance in this context based on the magnitude of $(C/f)\phi_0$, inspired by the corresponding distinction for bosonic parametric-resonance.} and used the WKB approximation, which produces results only for $k < (C/ f) \phi_0$, which is not a restriction for the analysis of Green and Kofman. However, the method of Peloso and Sorbo extends to the expanding universe case, as we see in later sections. The agreement between the two formulas is excellent for ${(C/ f)} \phi_0 > 1$ (the ``broad resonance'' regime) and there is an increasing difference as the coupling decreases (the ``narrow resonance'' regime). We tested these approximations against results obtained by solving the Dirac equation numerically on the homogeneous axion background for the broad resonance regime, where both formulae give identical results. Fig.\ \ref{fig:envelopebroad} shows that the agreement between the approximate and numerical results is exact at the points $t=2\pi n$ (where $n$ is an integer). That is, the approximation exactly reproduces the particle number after an integer number of inflaton oscillations. Fig.\ \ref{fig:envelope} shows the evolution of the fermion particle number for a system that is set at the threshold between broad and narrow resonance ${(C/ f)} \phi_0=1$.

\begin{figure}[t]
\centering
\includegraphics[width=3in]{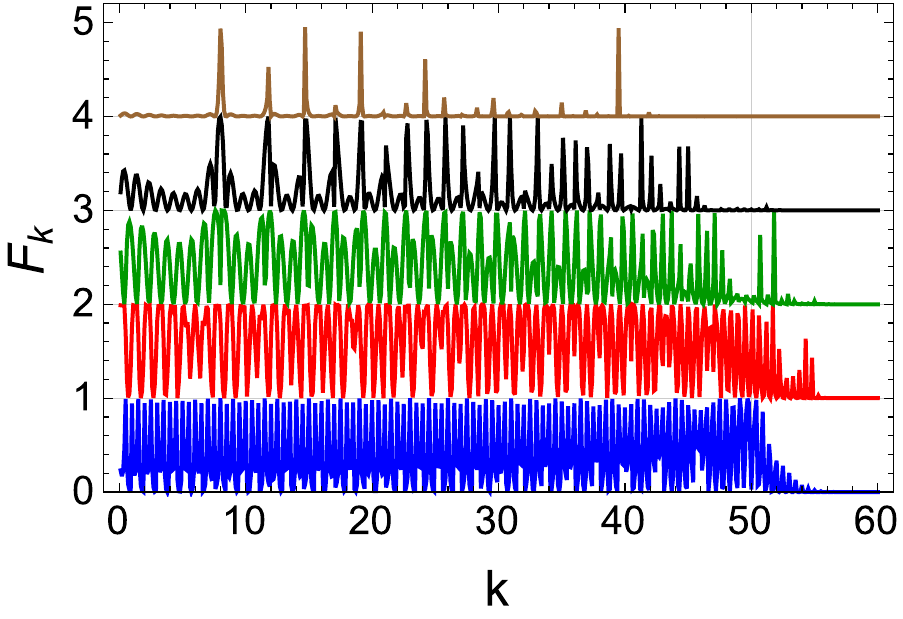} \includegraphics[width=3in]{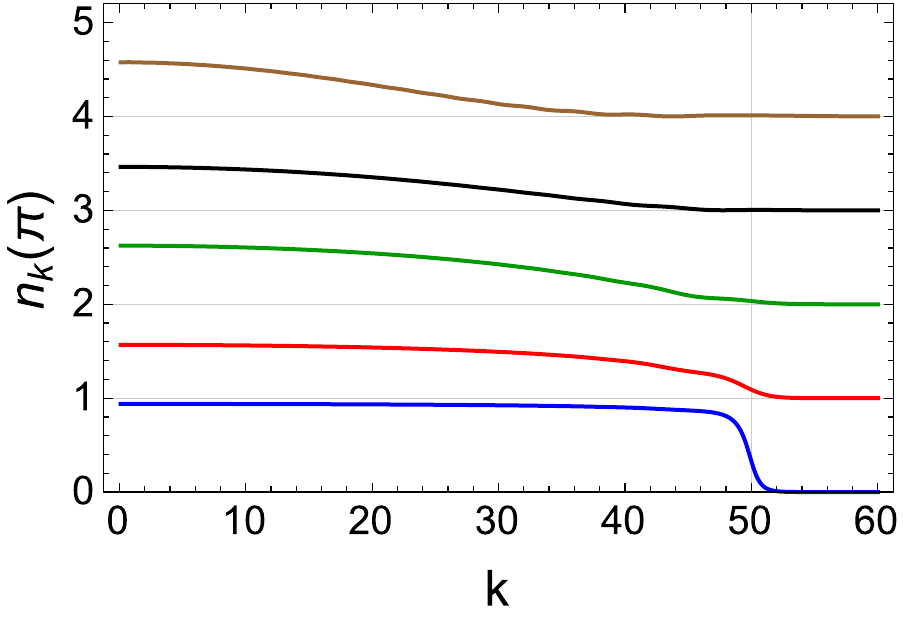}
\\
\includegraphics[width=3in]{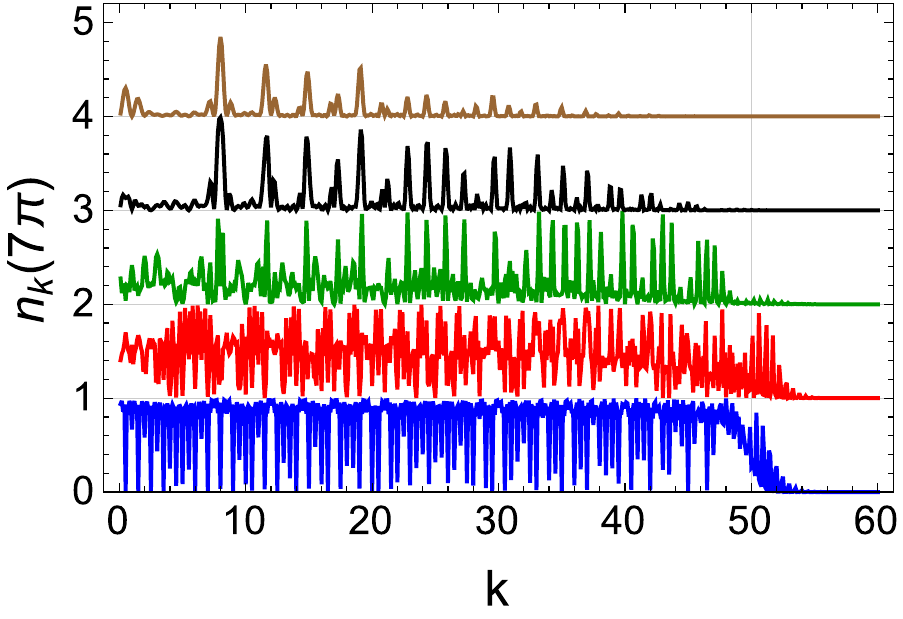} \includegraphics[width=3in]{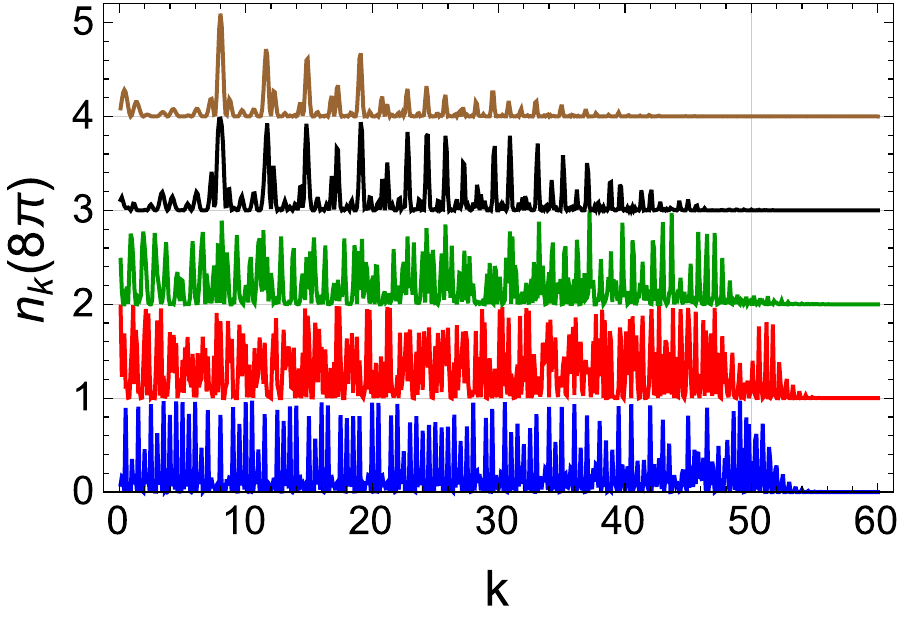}
\caption{ Particle number for $A=50$ and $m=1,3,5,7,10$ (blue, red, green, black and brown respectively). In the $n_k(\pi)$ plot the green curve is multiplied by $3$, the black curve is multiplied by $10$ and the brown curve is multiplied by $300$. In the $n_k(7\pi)$ and $n_k(8\pi)$ plots the brown curves are multiplied by $10$. All curves are shifted vertically by 1, 2, 3, and 4 for visual clarity. The vertical line at $k=A=50$ shows the maximum produced wavenumber according to the WKB method.}
\label{fig:A50varM}
\end{figure}

In Fig.~\ref{fig:A50varM} we show the results of numerically solving for the evolution of the particle spectra  for different fermion masses with $A = C/f\phi_0 = 50$,  chosen to be well within the region of validity of the WKB results. From comparison of the blue and red curves in the upper right panel of Fig.~\ref{fig:A50varM} with the corresponding curves in the lower panels, it appears that the mass  dependence of the final particle spectrum is weaker than that of the initial particle spectrum (the  particle spectrum after the first zero crossing of the effective wavenumber, $\tilde k =0$).  Further, for a range of masses, in our example $m \lesssim 5$, the particle spectra after an even number of zero-crossings of $\tilde{k}$ are more similar than their initial spectra would suggest. Lighter fermions are more easily produced in the first oscillations, yet the final occupation numbers appear to be similar, with the occupation number of several wavelengths reaching unity. This effect is due to the nature of fermion production. The occupation number of each mode $n_k$ cannot exceed unity, therefore systems with initial particle spectra of $n_k \sim 1$ can be either reduced or stay almost constant. This behavior causes the particle spectrum to develop fine band structure as parametric resonance develops, which can be averaged out  when calculating the total particle number. 

For light fermions with $m \lesssim 1$,  $n_k \approx 1$ for all excited modes after the first production event. In these cases, the second zero-crossing can only de-populate modes, since modes that are already at $n_k\approx 1$ cannot be further populated. This behavior results in  a decrease in the averaged particle occupation number after the second production event;  modes that were $n_k\approx 1$ after the first event end  up with occupation numbers $n_k \ll 1$ after the second event. This effect can be seen in Fig. \ref{fig:A50varM}, especially for the blue dotted curve.

Systems with heavier fermions have an initial particle production rate that is much less efficient, resulting in smaller initial particle spectra, $n_k \ll 1$. However, for these modes, each particle production event is more likely to create more particles than destroy them, resulting in an occupation number that grows with time. Given enough time these models can produce heavy fermions more effectively (at least for certain wavenumbers) than their initial production rate would suggest. In these cases, the final result is a particle spectrum that has broadly similar features to its lighter counterparts, as can be seen from the green curve in Fig.~\ref{fig:A50varM}. The important difference between the final particle spectra for fermions of different masses is in the range of wavenumbers that reach occupation numbers of $n_k\approx1$ rather then in the amplitude of the occupation number (which is $n_k \le 1$ by definition for fermions).

This picture changes dramatically once the expansion of the universe is taken into account. We show in the next section that the expansion introduces new effects which provide the means for generating a significant left-right helicity asymmetry in heavy fermions.
\begin{figure}[t]
\centering
\includegraphics[width=3in]{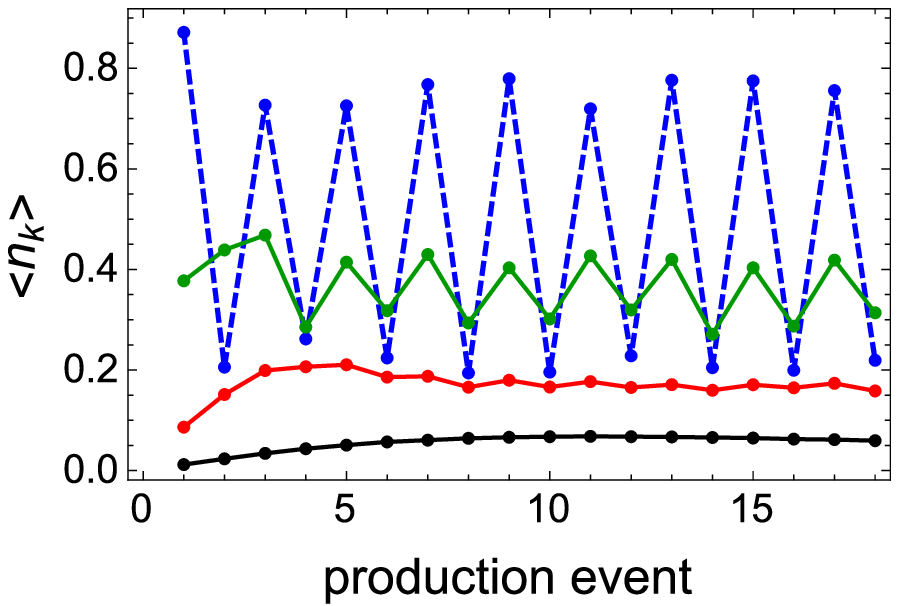}  \includegraphics[width=3in]{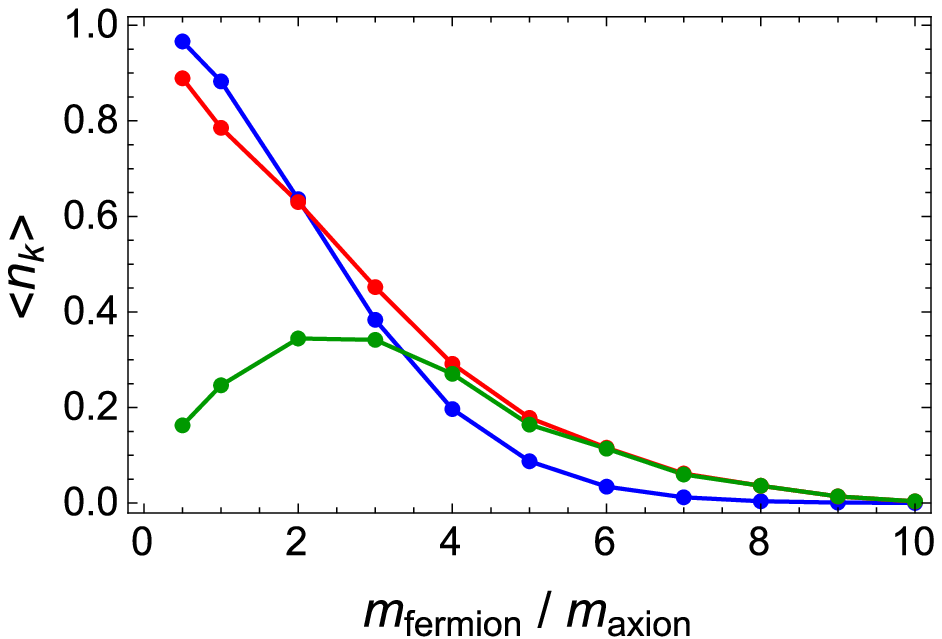}
\caption{
Left panel: The average occupation number as a function of the number of zero-crossings for $A=50$ and $m=1,3,5,7$ (blue-dotted, green, red and black respectively) in the static universe approximation. Right panel: The average occupation number as a function of the fermion mass. The blue curve corresponds to the first zero-crossing, while the red and green curves correspond to
$ \left < n_k (t=17 \pi)\right >$ and  $ \left < n_k(t=18\pi) \right >  $ respectively.}
\label{fig:navg_static}
\end{figure}
Before proceeding to the expanding universe case, we calculate the average occupation number per unit occupied volume in momentum space
\begin{align}
\left < n(t) \right > = { \int_0^{\infty}  n_k(t) k^2 {\rm d} k \over  \int_0^{k_{\rm max}}  k^2 {\rm d} k   }  = {3 \over k_{\rm max}^3}\int_0^{\infty}  n_k(t) k^2 {\rm d} k,
\label{eq:nAvgDef}
\end{align}
where  $k_{\rm max}$ is the maximum wavenumber that is excited according to the WKB approximation. For large values of the coupling, to a good approximation no particle production occurs for $k > k_{\rm max}$, and Eqn.\ \eqref{eq:nAvgDef} provides an estimate of the (weighted) average occupation number of each bin in $k$-space. We plot this quantity in Fig.~\ref{fig:navg_static} for $A=50$. However, we caution that this quantity is dependent on $k_{\rm max}$ itself. For small masses, where particle production is more effective, we see a large variation in the particle number after integer and half-integer numbers of axion oscillations. However, this is not true for larger masses where, as discussed above, after each axion-crossing the particle number is predominantly rising. In the expanding universe case,  this oscillating behavior of the average particle number is suppressed, since particle production after each production event is less efficient than the previous one due to the fact that the amplitude, and thus velocity of the axion oscillations, is damped by Hubble friction.

We end this section on the static universe approximation with a comment about the dependence of the maximum produced wavenumber on the coupling strength. According to the WKB approximation, particle production occurs only for ${k/ A} \le 1$. However, from Figs.\ \ref{fig:envelopebroad}, \ref{fig:envelope}, and \ref{fig:A50varM} note that this limit is increasingly violated as the coupling gets smaller. For couplings approaching unity there is significant particle production for ${k/ A}>1$, which is absent for larger couplings $A\gg1$. Parametric resonance allows for the production of these high-$k$ modes.


\subsection{Expanding universe}

The static universe approximation is a useful simplification that allows us to build intuition about the relevant physical processes without the complications of expanding space. However, away from the individual particle production events, it is not a good description of the processes occurring during and after inflation. Following inflation, the oscillation frequency of the background inflaton field is  typically not much faster than the expansion rate, and consequently it is important to take expansion into account.  As we demonstrate in this section, there are two main differences between the static- and expanding-universe cases. Both the range of wavenumbers that get excited and the particle production strength are progressively reduced by the expansion,  in contrast to the static universe, where they remain constant.

The main goal of this section is to estimate the final occupation number of the fermion states. Additionally, we calculate the initial and final fermion production asymmetry between the two helicity states. Since one helicity is being produced for the first time during inflation, while the other is produced only during preheating, we will demonstrate that unequal numbers of each helicity are produced in the overall evolution. We show that, because the behavior of both helicity states during the preheating phase is comparable and highly stochastic, the early behavior determines the ultimate fate of the asymmetry. Provided this initial asymmetry is large enough, we demonstrate that, at least in the model at hand, it can be preserved throughout the preheating period. This is quantified in Section \ref{sec:asym}.

Analogously to the WKB-based static universe approach, particle production in expanding spacetime occurs in distinct incidents at the times when $\tilde k (t_*)=0$, corresponding to 
\begin{align}
{k\over a (t_*)} \lambda = - {C\over f} \dot \phi (t_*).
\label{eq:zerocrossing}
\end{align}
Fermion production occurs in a very narrow region around $\tilde k =0$, therefore the expansion of the universe can be neglected during the production event and the static universe result of Eqn.\ \eqref{eq:nk_WKB} can be immediately applied in this case.  Different wavenumbers are excited at different times, according to Eqn.\ \eqref{eq:zerocrossing} and as shown schematically in Fig.\ \ref{fig:tildek}. The dependence of the particle production exponent on  helicity and  wavelength is hidden in the term
\begin{align}
\left |  {{\rm d} \tilde k \over {\rm d} t}\right |_{t_*}  = {C\over f} \left  | \dot \phi(t_*) H(t_*) +\ddot \phi(t_*) \right |, 
\label{eqn:dkdt}
\end{align}
where the time $t_*$ depends on $k$ through Eqn.\ \eqref{eq:zerocrossing}.

Despite describing particle production starting from a vacuum state, the form of $n_k$ given in Eqn.\ \eqref{eq:nk_WKB} is useful for calculating statistical properties of the particle spectrum, even after many production events and the emergence of a fine band structure, as shown in Fig.\ \ref{fig:A50varM} for the static universe case. We generalize this form in the following sections to include the case of $a(t) \ne {\rm const }$ and multiple productions. We consider $n_k$ as given in Eqn.\ \eqref{eq:nk_WKB} to be an indicator of the strength of any single particle production event. 

\subsubsection{During inflation}

As discussed earlier and in Section \ref{sec:inflation}, the axion-fermion coupling is active during the inflationary epoch. For one of the helicities  (in our case, with $\dot\phi < 0$ or equivalently $\vartheta >0$, the $\lambda = +1$ helicity) $\tilde{k}_\lambda$ may vanish \emph{during} inflation resulting in particle production during the inflationary epoch itself.  In this case, the time of the production event, where $\tilde k(t_*) =0$,  is found to be
\beq
t_* = {1\over H} \log \left ( { \vartheta H \over k \lambda } \right ) \, ,
\eeq
where we used the approximation of exact de-Sitter space ($H ={\rm const.}$) together with the slow-roll approximation $\ddot \phi =0$. We see that for $\dot \phi <0$ only the positive helicity state ($\lambda = 1$) can be excited during inflation.

The resulting fermion number with momentum $k$ can then be evaluated from  Eqn.\ \eqref{eq:nk_WKB} yielding
\beq
n_k = \exp \left  ( - {\pi m^2 \over |\vartheta| H^2} \right  ).
\label{eqn:nk_WKB_infl}
\eeq
We compare this result to the exact solution derived for particle production during inflation given in Eqn.\ \eqref{eqn:n_Whit}. Working in the limit  $\vartheta \gg 1$ (actually we do not need it to be much bigger than one, simply a few times larger), as well as ${m} = {\cal O}(H)$, Eqn.\ \eqref{eqn:n_Whit} becomes Eqn.\ \eqref{eqn:nk_WKB_infl} for the growing ($\lambda = +1$)  state. This WKB analysis agrees with the exact solution obtained for the modes during inflation, in the appropriate limit of large coupling (or $\vartheta \gg 1$), where the WKB method is applicable, leading to a unified treatment of the inflationary and post-inflationary fermion production for the parameter space of interest. It is worth noting that the WKB method employed here is valid only in the large-coupling regime, and therefore it cannot capture the particle production occurring solely due to the expanding space-time.

The mode that is excited at the end of inflation can be read off from Eqn.\ \eqref{eq:zerocrossing} to be $k = -{(C/ f)} \lambda \dot \phi(t_{\rm end}) a(t_{\rm end})$. In the case of chaotic inflation with a quadratic potential $ -\dot \phi(t_{\rm end}) a(t_{\rm end})\approx 0.7$.


\subsubsection{After inflation}
\label{sec:afterinflation}

\begin{figure}[t]
\centering
\includegraphics[width=5in]{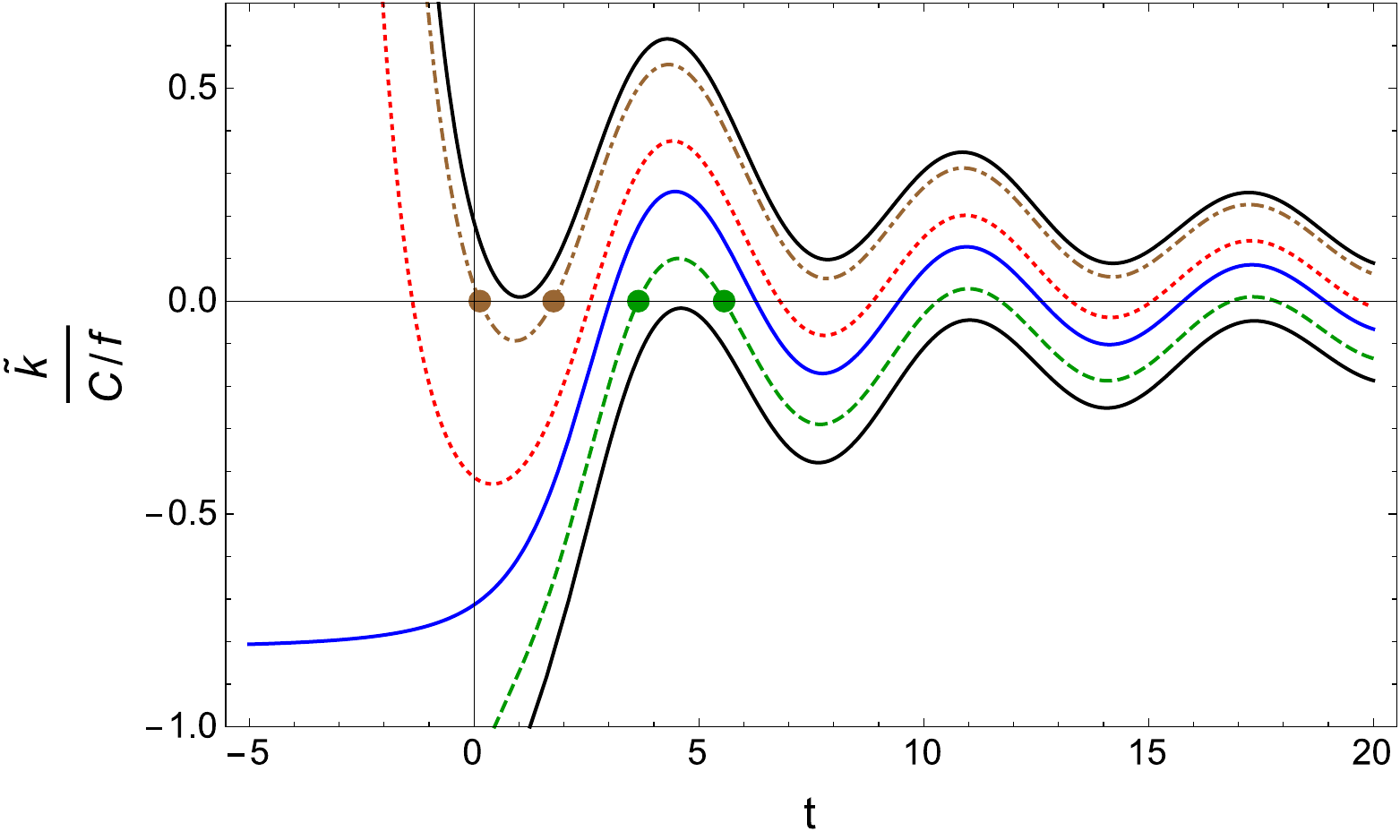} 
\caption{The effective wavenumber $\tilde k$ rescaled by the coupling $C/f$. The two black lines correspond to the maximum excited wavenumbers for $\lambda=-1$ (lower curve) and $\lambda=+1$ (upper curve). The blue line corresponds to $k=0$ and the green-dashed line corresponds to an intermediate curve for $\lambda=-1$. The red-dashed and brown-dot-dashed lines correspond to modes for $\lambda=1$ that  get excited for the first time during and after inflation respectively. Particle production occurs whenever $\tilde{k} $ crosses zero. Note that these particle production events occur in pairs which are closely spaced for high wavenumbers.}
\label{fig:tildek}
\end{figure}

Following the inflationary epoch, the axion oscillates about the minima of its potential. The particle production due to each zero-crossing of $\tilde k$ (neglecting the emerging band structure due to stochasticity) is governed by Eqn.\ \eqref{eq:nk_WKB} up to a maximum wavenumber given by
\begin{align}\label{eqn:kmaxafter}
k_{\rm max} ^\pm =\mp {C\over f} (a   \dot \phi )\big | _{\rm max},
\end{align}

We consider first the  $\lambda = -1$ helicity state which, for chaotic inflation with our conventions, is excited only after inflation has ended and the axion has crossed zero. The maximum particle number is produced for low wavenumbers, since $|dk/dt|_{t_*}$ is a decreasing function of $k$. For low wavenumbers, $k \ll C/f$, the condition $\tilde k =0$ becomes $\dot \phi (t_*)\approx 0$ and particle production occurs predominantly around points where the axion velocity vanishes. For chaotic inflation with a quadratic potential, $\ddot \phi(t_*) \approx 1/3$ during the first instance of  $\dot \phi (t_*)= 0$ and a simple application of Eqn.\ \eqref{eqn:dkdt} and Eqn.\ \eqref{eq:nk_WKB} leads to  
\beq
n_{k}\approx  \exp \left( - 9.4 {m^2 \over (C/f)  }    \right), \quad  k \le k_{\rm max}^- \approx 0.7 {C\over f}.
\eeq
 In the region of $C/f \gg 1$ the particle spectrum $n_k$ approaches a step function in wavenumber-space with a sharp cut-off at $k = k_{\rm max}^-$, hence the above formula, which suppresses the $k-$dependence, provides a useful estimate. Despite being a crude approximation, it provides an increasingly good estimate of the particle number for decreasing values of the combination $m^2 / (C/f)\ll 1$ and it can provide an upper limit on the particle production efficiency. 
 
We now  consider  the positive helicity state which becomes excited both during and after inflation.  The range of modes excited during inflation is $k \lesssim 0.7 {C/ f}$ for chaotic inflation, as shown in the previous section, while the total range of modes excited is $k \lesssim k_{\rm max}^+ \approx 0.9 {C/ f}$, obtained from Eqn.\ \eqref{eqn:kmaxafter}. The difference of $k_{\rm max}^+$ and $k_{\rm max}^-$ plays a key role in our discussion of asymmetry in Section \ref{sec:asym}.


An important characteristic of fermion preheating is the evolution of the Fermi sphere, defined in this case as the range of comoving wavenumbers that are (partially) filled with fermions as the universe expands during preheating. We again consider only the large coupling regime ${C/ f} \gg 1$, in order for the WKB approximation to be valid. For a matter dominated universe, arising from an axion potential that is predominantly quadratic during preheating, the inflaton amplitude decays as $\phi \sim { a^{-3/2}}$ and we may approximate
\begin{align}
\dot \phi  \sim \phi_0 {\sin(t + \delta \theta)\over a^{3/2}},
\end{align}
where $\delta \theta$ is an arbitrary constant phase. We then find a simple expression for the maximum excited wavenumber during a given production event
\begin{align}\label{eqn:kmax}
 k_{\rm max} = {1\over \sqrt a} {C\over f} \phi_0.
\end{align}
Since the scale factor increases,  the highest-momentum state that can be excited decreases as time progresses. For a locally quadratic minimum, the derivative of the effective wavenumber is given by
\begin{align}
\left | {{\rm d} \tilde k \over {\rm d} t} \right | =  {1\over a^{3/2}} \left | -{k\lambda\over a} + {C\over f} \phi_0  \right |  \approx {1\over a^{3/2}}{C\over f} \phi_0 \, ,
\end{align}
and thus the particle number can be written
\beq\label{eqn:productionspectrum}
n_k \sim \exp \left(- a^{3/2}\,\pi  { m^2  \over(C/ f) \phi_0} \right).
\eeq
Note that the particle production rate decreases as a function of time. When taken together with Eqn.\ \eqref{eqn:kmax}, Eqn.\ \eqref{eqn:productionspectrum} implies the first few production events are the most important for determining the characteristics of the particle spectrum produced in this model. These events produce the most particles out to the largest wavenumber with the greatest efficiency. Subsequent production events can only affect smaller and smaller wavenumbers and with a decreasing strength. These subsequent events simply introduce a band structure that can be averaged out for all practical purposes, analogously to the static universe case.

Finally, we note the difference between the case considered here and the case of Yukawa-coupled light fermions ($m \ll m_\phi$). In the limit of large Yukawa coupling,  the maximum comoving wavenumber that is excited grows with the expansion as $k_{\rm max} \sim a^{1/4}$ where $a(t)$ is the scale-factor, normalized as $a=1$ at the end of inflation  \cite{Greene:2000ew}. This result can be derived by noting that the particle production probability is given by
\begin{align}
n_k \sim \exp\left( -\pi{ ({k/ a})^2 \over  {h \phi_0 / a^{3/2}} }\right),
\end{align}
and thus the time dependence of the maximum wavenumber excited can be read off as  $k_{\rm max} \sim a^{1/4}$. In the case of heavy fermions ($m > m_\phi$) the result is more complicated, giving the final form of $n_k$ as a shifted Gaussian in $k$
 (when studied numerically, this Gaussian is found to contain even more structure), rather than the smooth fermi sphere of the light fermion case. 
\begin{figure}
\centering
\includegraphics[width=3in]{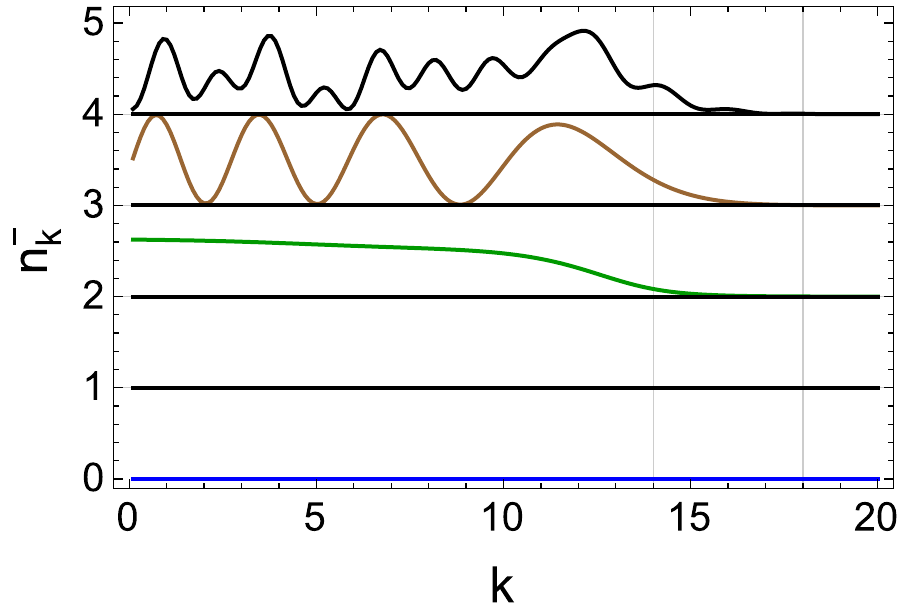} \includegraphics[width=3in]{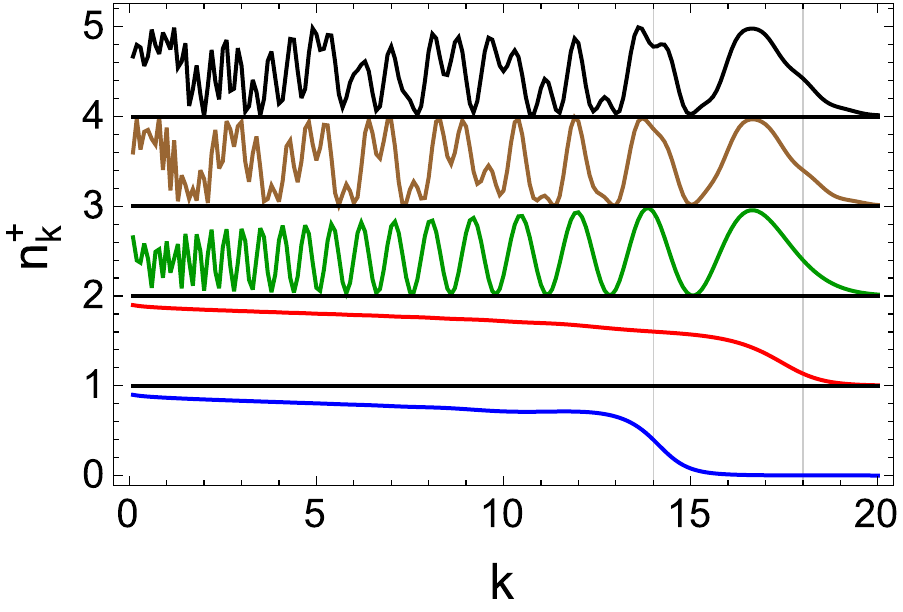}
\caption{Particle number for $m=1$, $C/f=20$ and $\lambda=-1,1$ (left and right respectively) in an expanding universe.  The different curves show the particle number at $t=0$, at the time when all $k$-modes of the $\lambda=1$ state have been produced for the first time and after $1,2,3$ zero-crossings respectively from bottom to top. Curves are shifted vertically for clarity. The vertical lines correspond to $k=0.7C/f$ and $k=0.9C/f$.}
\label{expand_s1sm1}
\end{figure}

As time progresses, the successive production events at each point where $\tilde k=0$  change the particle number with decreasing effect. This is illustrated in the upper panels of Fig.\ \ref{fig:nav_num} where we show the average particle occupation number of Eqn.\  \eqref{eq:nAvgDef} as a function of time. We have so far only considered particle production from the first time where $\tilde{k} = 0$. In order to accurately estimate the final number densities for each helicity state, we extend our analysis to successive events. 

The problem of several particle production events can be mapped to the well-known quantum-mechanical problem of a particle undergoing multiple scatterings. In the static universe the constructive or destructive interference creates a well-defined band structure. However, in the expanding universe case, successive phases can be thought of as random. The calculation is analogous to that in  \cite{Peloso:2000hy}, and we simply quote the result here, referring the reader to the original works for more details. After $N$ successive production events the smoothed particle occupation number is given by
\begin{align}
n_k^{(N)}    = {1\over 2} - {1\over 2} \prod_{i=1} ^N \left (1-2e^{-\pi p_i^2} \right ) \,.
\label{eqn:nkN}
\end{align}
This result shows that Eqn.\ \eqref{eq:nk_WKB} can be used as to characterize particle production at all times, despite being strictly true only for the first particle production event. We test its accuracy in the next section. 

Fig.\ \ref{fig:nav_num} shows the average particle number as a function of the successive $\tilde k$ zero-crossings for a fixed value of the coupling ${C/ f} = 50$ and varying fermion mass. The first result is the fact that particle production is suppressed for larger values of the mass, and effectively shuts off for ${m^2 /( C/f)} \gtrsim 0.5$, and ${m^2 / (C/f)} \gtrsim 0.3$, equivalently $m\gtrsim 5$ and $m \gtrsim 4$,
for the positive and negative helicity mode respectively. A similar behavior was seen in Fig.~\ref{fig:navg_static} for the static universe analysis.

The second important feature of Fig.\ \ref{fig:nav_num} is the decline of the final particle number for small values of the mass. This presents a major departure from  the results of the static universe analysis. In the static universe approximation, the particle number oscillates between two extrema for small masses, as shown in Fig.\ \ref{fig:navg_static}.  From Fig. \ref{fig:tildek}, note that particle production events always come in pairs of similar strength, especially for wavenumbers near $k_{\rm max}$. This is shown in Fig.\ \ref{fig:tildek} by the two pairs of brown and green dots for the $\lambda =+1$ and $\lambda=-1$ states respectively, which represent successive production events. For lighter fermions,  particle production is very efficient, and the particles that are created by the first event tend to be destroyed by the second event. In the static universe the production efficiency is constant, but in the expanding universe subsequent events are  much less efficient, as seen from Eqn.~\eqref{eqn:productionspectrum}. Furthermore, as show in Section \ref{sec:afterinflation}, particles are produced over a smaller range of wavenumbers as the universe expands. Taken together, somewhat surprisingly, the net result is a low overall particle number for light fermions. We also demonstrate this result  analytically. Applying Eqn.\ \eqref{eqn:nkN} and taking the first two particle production events to be very efficient, that is $e^{-\pi p_i}=1-\epsilon_i \approx 1$, we get 
\begin{align}
 n_k^{(2)}= \epsilon_1+\epsilon_2-2\epsilon_1 \epsilon_2 \approx   \epsilon_1+\epsilon_2 \ll1 \, .
\end{align}
meaning that (on average) modes that start almost fully occupied become almost empty after the second production event.

An important qualitative feature of the expanding universe case is the similarity between the particle spectra after two production events, $  n_k^{(2)}$, and the result after a very long time, $ n_k^{(\infty)} $. Successive production events become increasingly less efficient and affect a decreasing range of wavenumbers, therefore these first two production events give the dominant contribution to the final particle spectra. We have already seen in the case of light fermions that these two events lead to the creation of a large number of particles followed by the immediate destruction of many of them. The upper panels of Fig.\ \ref{fig:nav_num} show that particle number is effectively constant after the first two oscillations. This is important because it means that any helicity asymmetry will be generated and finalized almost immediately after the end of inflation.

Production events can be always described in pairs, therefore simple intuition tells us that, on average, modes with $n_k <0.5$ initially tend to grow their population towards unity after the second $\tilde k=0$ point, while on average, modes with $n_k >0.5$ initially tend to reduce their population (away from unity). At fixed coupling $C/f$, production of lower mass fermions is more efficient, and they are mostly destroyed by the double production event. In the large mass region on the other hand, the efficiency of each event is lower, but successive scatterings are more likely to be constructive. Taken together, this suggests that the first two production events will produce a maximum of $ \left < n_k \right >\approx 0.5$. This is exactly what the blue and red lines in the lower panels of Fig.\ \ref{fig:nav_num} show.


\subsubsection{Asymmetric helicity production}
\label{sec:asym}

From our discussion of the range of excited wavenumbers and the evolution of the Fermi sphere, we already see a first hint of helicity asymmetry, simply from the fact that the positive helicity mode is produced up to a larger wavenumber $k_{\rm max}^+ > k_{\rm max}^-$. 
With the average particle number  in Eqn.\ \eqref{eq:nAvgDef} and the total particle number density
\begin{align}
n = \int {\rm d}^3k\, n_k,
\end{align}
the particle asymmetry between the two helicity states becomes
\begin{align}
\nn 
\Delta n=  n ^+   -  n^-   &\approx   4\pi \int_0^{k_{\rm max}^-} k^2\left (n_k^+  - n_k^-\right ) {\rm d}k  + 4\pi \int _{k_{\rm max}^-} ^ {k_{\rm max}^+} k^2 n_k^+ {\rm d} k 
\\
&= {4\pi\over 3}(k^-_{\rm max})^3 \left < n^+ - n^- \right >_0^{k^-_{\rm max}}  +
{4\pi\over 3} \left ( (k^+_{\rm max})^3 -  (k^-_{\rm max})^3 \right )  \left < n^+ \right >_{k^-_{\rm max}} ^{k^+_{\rm max}} .
\label{eqn:navplus_navminus}
\end{align}
For $m_\phi^2\phi^2$ inflation, the first term scales as $(k^-_{\rm max})^3 \approx 0.7^3 \left ({C/ f}\right )^3 \approx 0.34 \left ({C/ f}\right ) ^3 $, while the second term scales as $ \left ( (k^+_{\rm max})^3 -  (k^-_{\rm max})^3 \right )  \approx (0.9^3 - 0.7^3) \left ( {C/ f}\right )^3 \approx 0.39 \left ( {C/ f}\right )^3$. The two terms can therefore be comparable, depending on the structure and values of $n_k^\pm$.  Having given estimates of the range of excited wavenumbers, we can now estimate the particle number and study the cases where it is possible to generate large asymmetries in the different fermion helicities.
\begin{figure}[t]
\centering
\includegraphics[width = 3in]{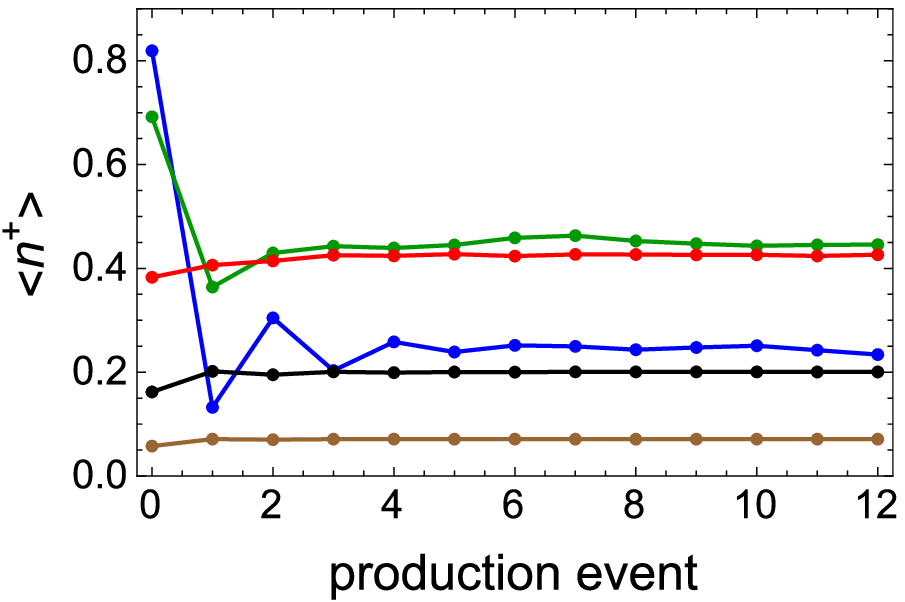} \includegraphics[width = 3in]{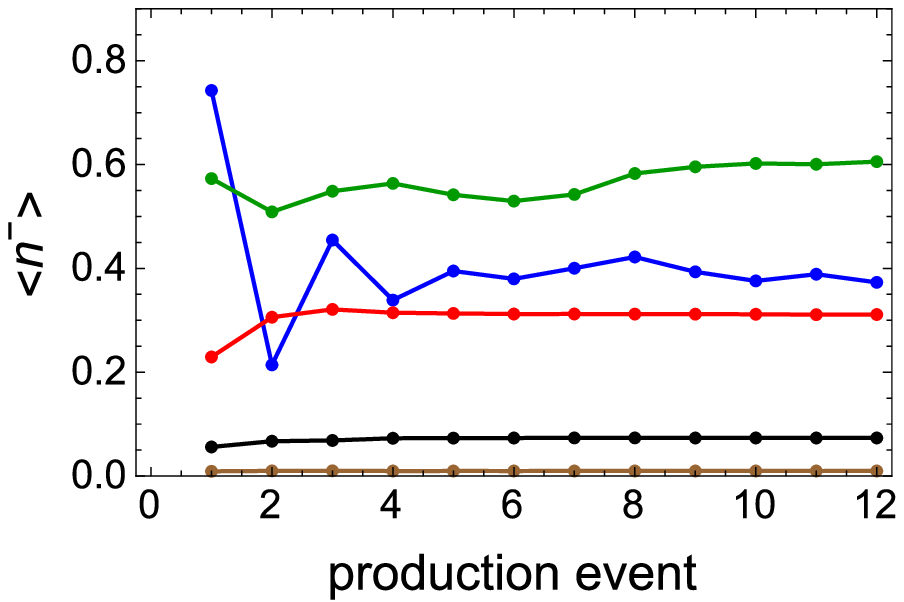}
\\
\includegraphics[width = 3in]{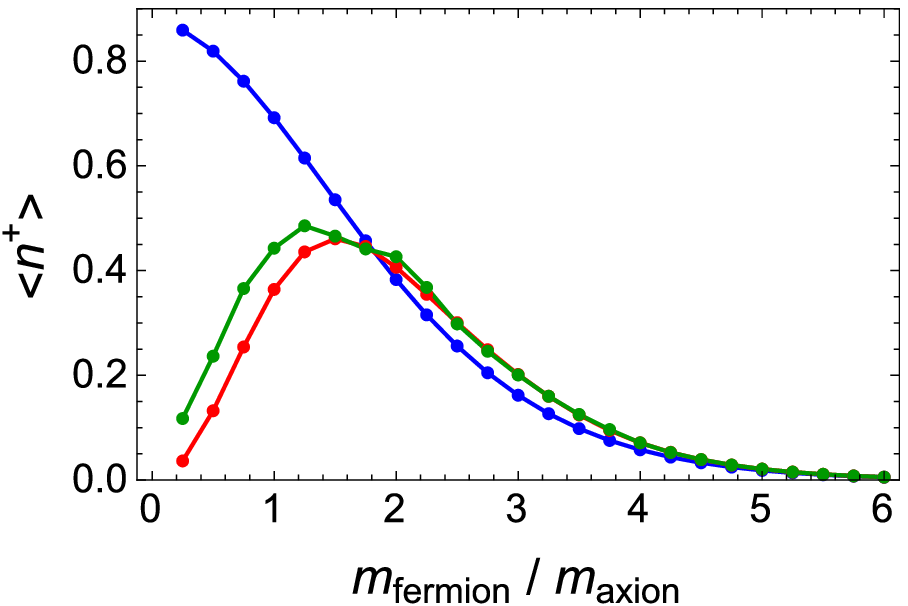} \includegraphics[width = 3in]{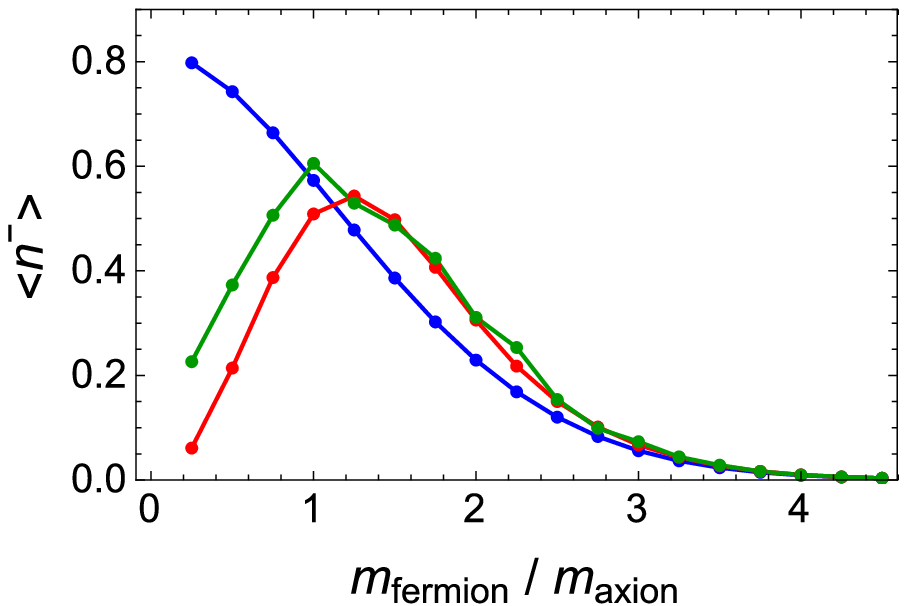}
\caption{Upper panels: Average particle number after successive production events for the positive (left)  and negative (right) helicity states for ${C/ f} = 50$ and $m_{\rm fermion} = 0.5,1,2,3,4$ (blue, green, red, black and brown respectively). The $0$'th production event occurs during inflation.
Lower panels: Average particle number after the first (blue) and second (red) production event and after multiple $\tilde k$ zero-crossings (green).
 }
\label{fig:nav_num}
\end{figure}

Returning to Fig.\ \ref{fig:nav_num}, we note that positive and negative helicity states have comparable average particle occupation numbers. However, when converting this to a total particle number, one has to take into account the fact that each helicity state occupies a different volume in phase space. For $m_\phi^2 \phi^2$ inflation, for example, $k^+_{\rm max} = 0.9 {C/ f}$ for the positive helicity state and $k^-_{\rm max} = 0.7 {C/ f}$ for the negative helicity state, which means that the particle number of the positive helicity state is more than twice that of the negative helicity state for $\left <n^+ \right > = \left <n^-\right >$. Thus  the positive helicity mode is dominant, and a helicity asymmetry is present. We will quantify this asymmetry further using numerical simulations.

\subsubsection{Numerical results}

To test our analytic and semi-analytic results, we numerically evolve the equations of motion for the fermions in the homogeneous axion and gravitational field. We evaluate the evolution of the fermion number for a range of couplings $2\le M_{\rm Pl} C/ f \le 500$, and fermion masses, such that $ 0< {m^2 / (C/f)} < 1$. We neglect the back-reaction of the produced fermions on the axion evolution; back-reaction can be safely ignored for values of the coupling that satisfy $(C/ f)M_{\rm Pl} <10^3$, at least during inflation and for the first stages of preheating, where all interesting effects are expected to occur.

Later axion oscillations (as shown in  Fig.\ \ref{fig:nav_num}) become increasingly irrelevant, and so we focus on the first couple of fermion production events, which determine the final result to a good approximation. For the first two production events Eqn.\ \eqref{eqn:nkN} becomes for the two helicity states
\begin{align}
\nn
n_k^+ \approx {1\over 2} - {1\over 2}  \left (1-2e^{-\pi (p_0^+)^2} \right ) \left (1-2e^{-\pi (p_1^+)^2} \right ),
\\
n_k^- \approx {1\over 2} - {1\over 2}  \left (1-2e^{-\pi (p_1^-)^2} \right ) \left (1-2e^{-\pi (p_2^-)^2} \right ),
\label{Eq:WKBsummed}
\end{align}
which provide an increasingly good approximation for the numerically calculated result for $C/f \gg1$, both after the second production event as well as after multiple ones. 

\begin{figure}[h]
\centering
\includegraphics[width =3in]{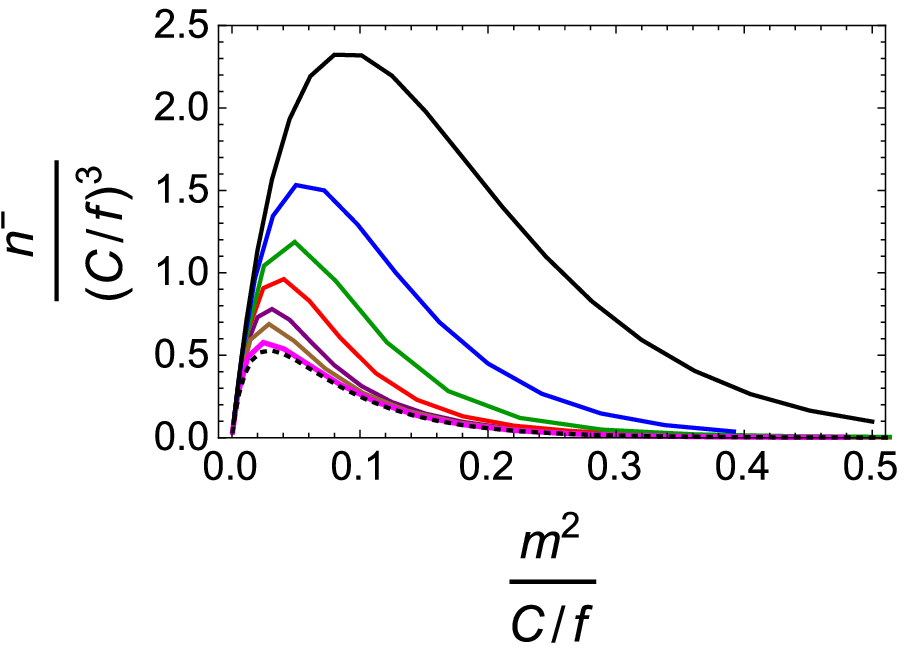} \includegraphics[width =3in]{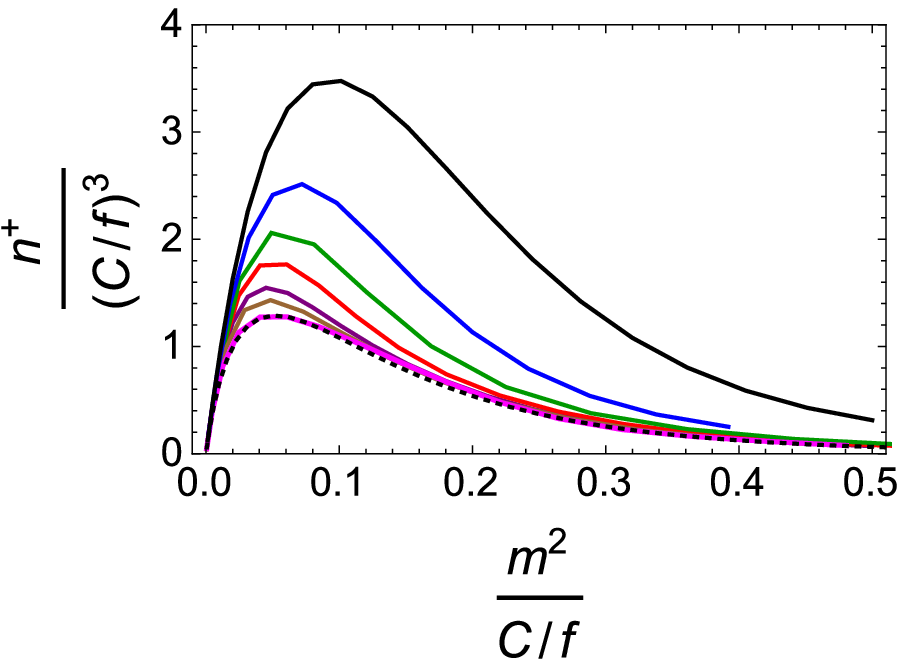}
\caption{Particle number density after two production events for the negative (left) and positive (right) helicity state for chaotic inflation.  Different colors correspond to different values of the axion-fermion coupling ${C/ f} = 2, 5, 10, 20, 50, 100, 500$ from top to bottom.  The black-dotted line corresponds to the WKB result given by Eqn.~\ref{Eq:WKBsummed}.   }
\label{fig:Nplusminus}
\end{figure}

The important feature of the WKB-based results is their dependence solely on the combination $m^2/ (C/f)$, rather than on the fermion mass and coupling separately. However, as shown in Fig.~\ref{fig:Nplusminus}, this is  not strictly true for the particle number obtained from a numerical evolution of the system. There is a significant departure from the WKB result that occurs for wavenumbers that are near the maximum excited wavenumber. In particular, smaller couplings tend to have a larger ``tail'' in the $n_k$ distribution. This is  the same behavior manifest in the static universe case. Because there is much more phase space at high wave number, the volume factor ${\rm d}^3k $ weighs these modes more, and this discrepancy widens when the total particle number density or average particle occupation number $\langle n_k \rangle$ is evaluated. This phenomenon occurs for the positive and negative helicity states alike, leading to an enhanced  $\left < n_k \right >$ at smaller values of the couplings $C/f$, for the same ratio $m^2 / (C/f)$. 

For order unity values of the coupling, the departure from the semi-analytic result is even larger -- this is perhaps not surprising as the WKB approximation breaks down in this region. After several production events, these smaller couplings exhibit a slow but steady parametric excitation of larger wavenumbers, reaching $k_{\rm max} \sim 2 {C/ f}$. Despite the occupation number being small $n_k \ll 1$ in this range, the particle number is again enhanced due to the ${\rm d}^3k$ phase-space factor. Because we calculate the total particle density by integrating $n_k$ up to a sufficiently large wavenumber in order to enclose all particle production, and normalize it by $k_{\rm max}^3$ (given by the WKB approximation), the average particle number density $\langle  n^\pm \rangle$ can exceed unity for late times and small couplings. Furthermore, because of the steady excitation of these short-wavelength modes, the particle number keeps rising as time progresses, and there is no well-defined asymptotic particle number for late times, contrary to the case of larger couplings. A further effect of the behavior of these modes is the restoration of the symmetry between negative and positive helicity states. This leads to $\Delta n = n^+ - n^-$ taking both positive and negative values for ${C/ f} =2$ in our simulations. However, the range of ${C/ f} ={\cal O}(1)$ is  not of particular interest. This is due to the fact that the total number of particles (and the corresponding helicity asymmetry) is proportional to the cube of the maximum wave number $k_{\rm max}^3 \propto \left ({C/ f} \right )^3$, which implies that for a considerable effect to be achieved, we need ${C/ f} \gg1$. This makes model-building more robust, since ${\Delta n/ (C/f)^3}$ is largely insensitive to the value of the coupling in this region.

\begin{figure}[h]
\centering
\includegraphics[width =3in]{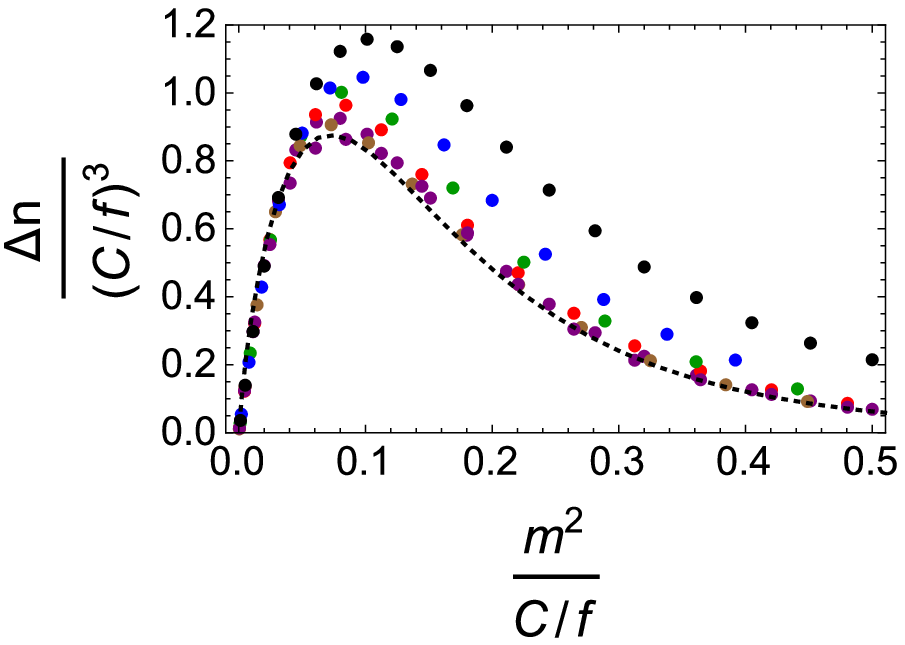} \includegraphics[width =3in]{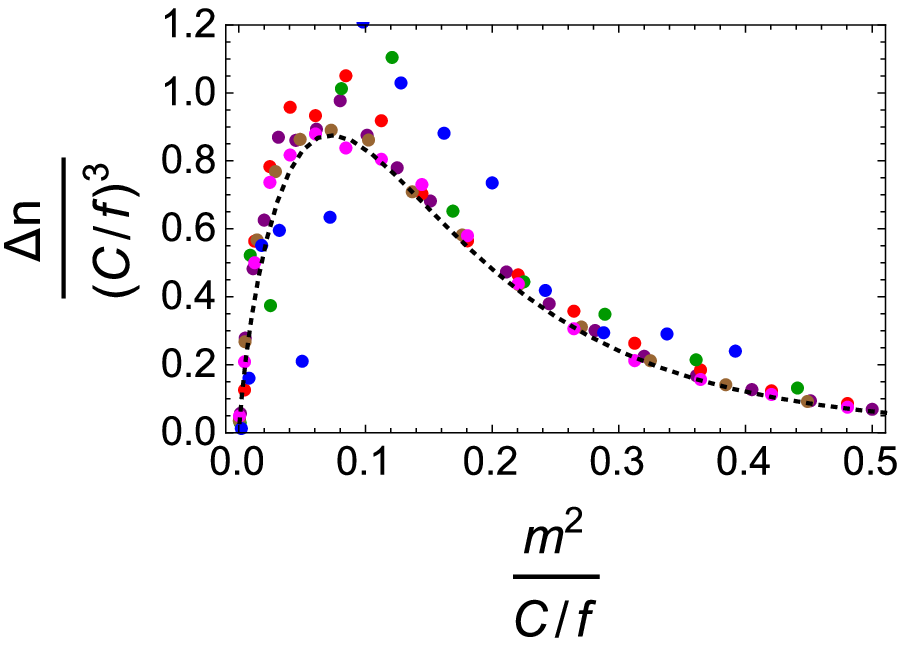}
\caption{ Measure of the total helicity asymmetry $\Delta n = n^{(+)} - n^{(-)} $ rescaled by the axion-fermion coupling cubed $(C/f)^3$ after the second axion crossing (left) and after many ($10$) axion crossings (right) for chaotic inflation.  Different colors correspond to different values of the axion-fermion coupling, consistent with Fig.~\ref{fig:Nplusminus}, while the black-dotted line corresponds to the WKB result of Fig.\ \ref{fig:Nplusminus}. }
\label{fig:asym_sim}
\end{figure}

Fig.~\ref{fig:asym_sim} shows the total particle number density asymmetry in helicity states $\Delta n $. This quantity scales with the cube of the coupling, so we have divided out by a factor of $( C/f )^3$ to put all curves on the same axes. The main results of this calculation are the dependence of the normalized helicity asymmetry $\Delta n / (C/f)^3$ on the number of axion oscillations, as well as on the coupling $C/f$ and the fermion mass $m$, or equivalently on one of these two parameters and the ratio $m^2 / (C/f)$. 
Fig.~\ref{fig:asym_sim} illustrates that, for large values of the coupling $C/f \gg 1$, the normalized helicity asymmetry is largely insensitive to the value of the coupling and is only determined by the ratio $m^2/( C/f)$. Furthermore, there is no significant evolution of the normalized helicity asymmetry after the second production event. We omit the case $C/f=2$ from the right panel of Fig.\ \ref{fig:asym_sim}, due to the absence of a well defined late-time asymmetry, as explained above.

From a model-building perspective, for a fixed fermion mass $m$, the resulting asymmetry can be easily read from Fig.~\ref{fig:asym_sim} for different values of the coupling. The maximum value of the asymmetry occurs for $m^2 / (C/f) \approx 0.07$ and the full width at half maximum is given by 
\begin{align}
0.02 \lesssim {m^2\over C/f}  \lesssim 0.2 \, ,
\end{align}
which serves as a condition for the existence of a strong helicity-asymmetry. In order to provide a rough, but intuitive estimate of the maximum asymmetry strength, we start with the observation that the average occupation number for each helicity state,  when considered as a function of $m^2/(C/f)$,
 peaks at $\left < n_k^{\pm} \right >_{\rm max} \approx 0.5$. The total particle number is
\begin{align}
n^ \pm = \int {\rm d}^3 k\, n_k = 4\pi {(k_{\max}^\pm )^3 \over 3} \left < n^\pm \right > \approx  {2\pi\over 3} (k_{\max}^\pm )^3 ,
\end {align} 
 where we used the peak value $\left < n \right > \approx 0.5$. By assuming that the two distributions for the $n_k^+$ and $n_k^-$ peak at the same value of $m^2 / (C/f)$ the asymmetry is simply
 \begin{align}
 \Delta n =   {2\pi\over 3} \left [ \left ({k_{\max}^+ \over C/f }\right )^3 - \left ({k_{\max}^- \over C/f }\right )^3 \right ] \left ( {C\over f}\right ) ^3.
 \label{eq:asym}
 \end{align}
Thus, for the peak value of the asymmetry, we see that the first term in Eqn.\ \eqref{eqn:navplus_navminus} vanishes, while the particle number density average in the second term can be taken as approximately $0.5$.
 For chaotic inflation with a quadratic potential, the values of $k_{\max}^\pm$ lead to
 \begin{align}
 \Delta n_{\rm m^2\phi^2} \approx 0.8  \left ({C\over f} \right )^2 \, .
 \end{align}
 This formula provides a slight underestimation of the peak helicity asymmetry by about $10\%$, as seen in Fig. \ref{fig:asym_sim}. However, both the accuracy, as well as the simplicity of this formula make it very useful for exploring the fermion production capability of different inflationary models, like the axion monodromy potential which will be discussed in the next section.

\section{Axion Monodromy Inflation}

In this section, we extend the results derived above in Section \ref{sec:quadinf} for quadratic chaotic inflation to the simplest model of axion monodromy inflation with a potential 
 given by Eqn.\ \eqref{eqn:monopot}. All the machinery we have developed in Section \ref{sec:quadinf} can be transferred directly to this case. The only differences are in the specific numerical values of $ | {d\tilde k / dt} |$ at the various production events and the range of excited wavenumbers. 

We begin by rescaling our parameters to rewrite the background inflaton equation of motion in a dimensionless form. To generate a spectrum of curvature fluctuations with an amplitude that matches the observed microwave background, we choose $\log_{10}(\mu/M_{\rm Pl})  = -3.22$ \cite{Peiris:2013opa}. Measuring the inflaton field in units of the Planck mass, and time in units of $1/\sqrt{  M_{\rm Pl} / \mu^3}$, the background equation of motion for the inflaton becomes
\begin{align}
   \ddot \phi + 3 H  \dot \phi + {  \phi \over \sqrt{  \phi^2 + \phi_c^2}}=0  \, .
\end{align}
In what follows, wavenumbers will be measured in terms of $ \sqrt{\mu^3 / M_{\rm Pl}}$.

Most of the qualitative features of the $m_\phi^2\phi^2$ case can be directly applied to monodromy inflation. For monodomy inflation, gravitational particle production during inflation is biased by the rolling axion so that only a single helicity is produced. After inflation both helicity states are excited at the points where $\tilde k(t)=0$, which differ for each helicity state and wavenumber. The range of excited wavenumbers is smaller for each subsequent production event, due to the decay of the inflaton oscillations, making the first two production events the dominant ones. For axion monodromy inflation, there is an additional parameter, $\phi_c$, which controls the shape of the potential near the end of inflation and quantitatively changes the results.

\begin{figure}[h]
\centering
\includegraphics[width =2.95in]{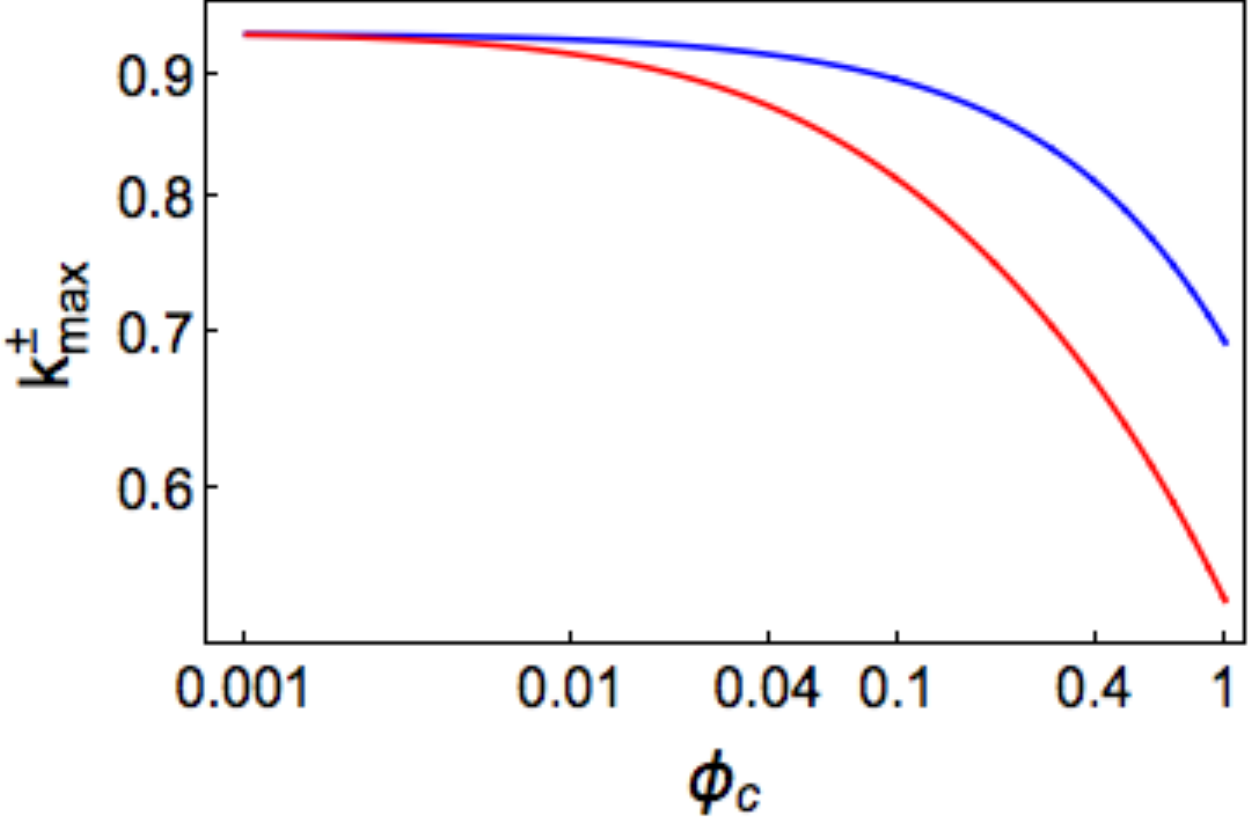} \includegraphics[width =3.05in]{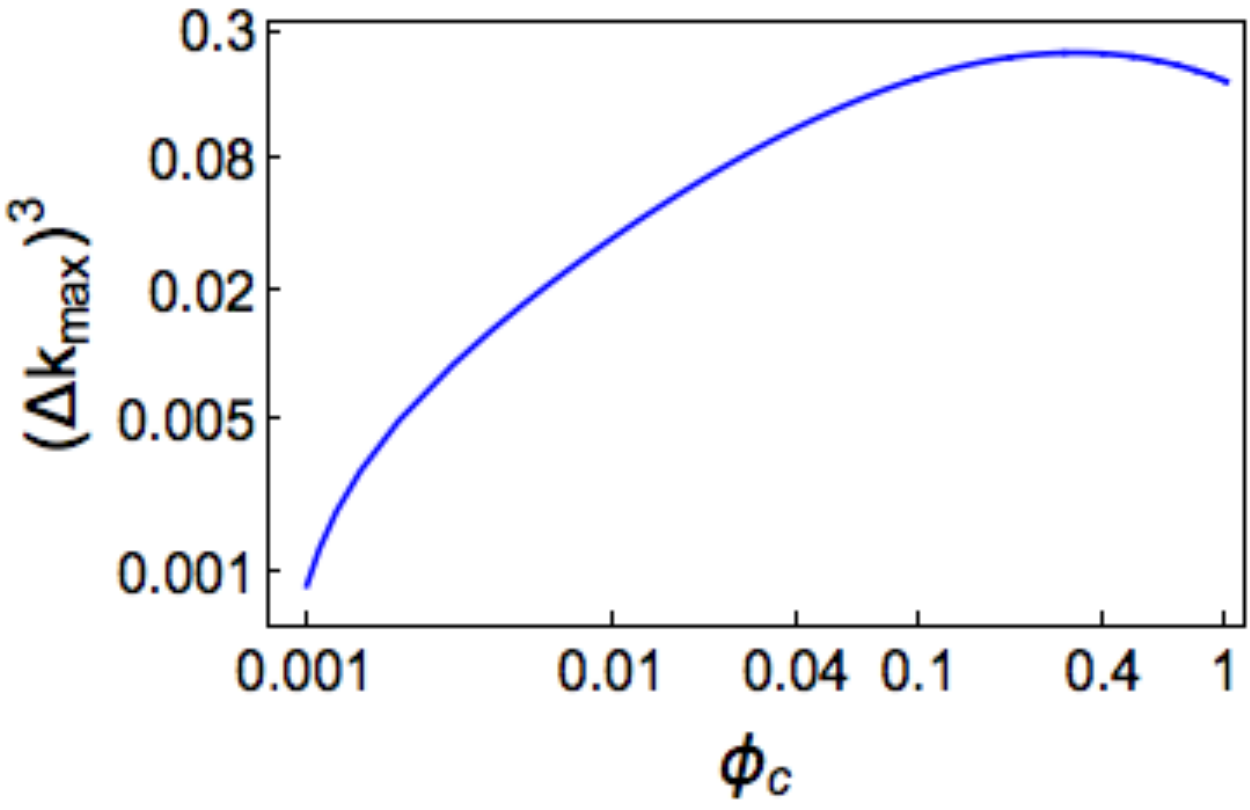}
\caption{Left panel: The maximum initially excited wavenumber for $\lambda=+1$ (upper blue curve) and $\lambda=-1$ (lower red curve) is shown in units of $C/f$ for axion-monodromy inflation. Right panel: Plotted is the ratio $ \left [ \left (k_{\rm max}^+ \right )   - \left (k_{\rm max}^- \right )   ^3\right ]/( C/f)^{3}$, which can be used to estimate the helicity asymmetry from Eqn.\ \eqref{eq:asym}.}
\label{fig:kmaxMono}
\end{figure}

 For $\phi_c \ne 0$ the first few oscillations dominate and define the range of the Fermi sphere, as we found in the $m_\phi^2\phi^2$ case. The corresponding maximum wavenumbers excited during the first production event for each of the two helicity states are shown in Fig.\ \ref{fig:kmaxMono}, along with
\begin{align}
(\Delta k_{\rm max})^3 \equiv {  \left (k_{\rm max}^+ \right )   - \left (k_{\rm max}^- \right )   ^3 \over (C/f)^3} \, ,
\end{align} 
which tracks the difference in phase space volume that is populated in each case. Note that for $m^2\phi^2$ inflation $(\Delta k_{\rm max})^3 \approx 0.4$, which is similar to the values for axion monodromy inflation, in the range of $0.1< \phi_c<1$.  

For large values of $\phi_c\geq$, the minimum of the potential behaves more and more like a quadratic model during the oscillation phase, and we find similar quantitative results to $m_\phi^2\phi^2$ inflation. Specifically for $\phi_c=1$ the form of $a(t) \dot \phi(t)$ is very similar for quadratic and axion monodromy inflation, with a difference in amplitude of about $25\%$. This leads to a difference in the number density of produced fermions of about $1.25^3 \approx 2$.
The maximum wavenumber excited in each axion oscillation, $k_{\rm max}^{\pm}$, grows with decreasing $\phi_c$, as shown in Fig. \ref{fig:kmaxMono}. However, the combination $(k_{\rm max}^{+})^3- (k_{\rm max}^{-})^3$ decreases for $\phi_c \lesssim 0.3$. For small values of $\phi_c$, both helicities are produced in increasingly equal amounts, resulting in smaller helicity asymmetries. The reason for this behavior can be seen from the (unphysical) limit  $\phi_c \to 0$. In this limit, $\dot \phi (t)a(t)$ oscillates with a constant amplitude generating a Fermi sphere with a constant radius in momentum space. Each axion oscillation therefore produces particles out to the same maximum wavenumber with an efficiency that remains essentially constant. Hence for $\phi_c <0.1$ we have an increased particle production (due to the increased range of produced wavenumbers) and a decreased level of helicity asymmetry. 

\begin{figure}[h]
\centering
\includegraphics[width =3in]{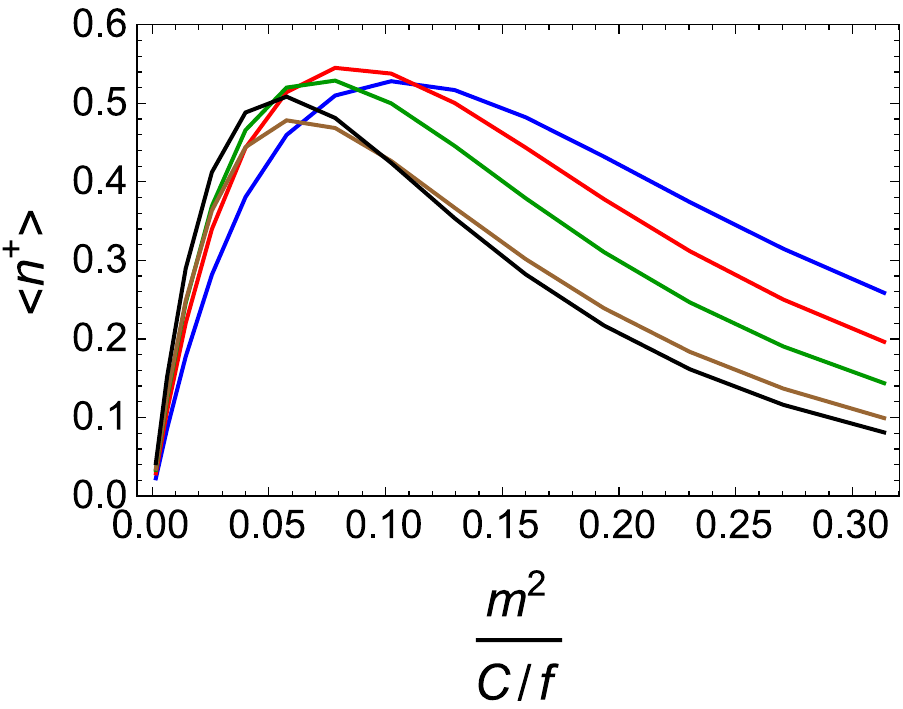} \includegraphics[width =3in]{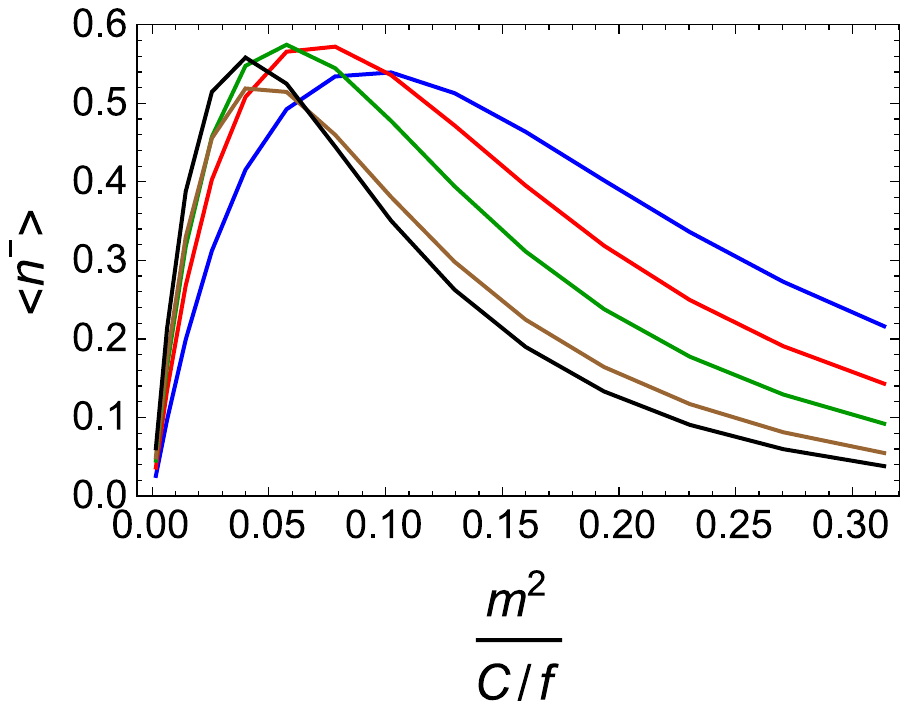}
\caption{Average particle number after two production events for the case of axion monodromy inflation for $\phi_c = 0.2, 0.3,0.4,0.5,0.6$ (blue, red, green, brown and black respectively). The maximum value of $\left < n^\pm \right > \approx 0.5$ is evident.}
\label{fig:nMono}
\end{figure}

The average particle number after the second production event for each helicity state is shown in Fig.\ \ref{fig:nMono}. This is a very good approximation for the final particle number, especially near the peak of the curve $\langle n ^\pm \rangle$ as a function of $m^2/(C/f)$, where both the maximum particle production and the maximum asymmetry occur. We present results for the range $0.2<\phi_c<0.6$, where the asymmetry is largest, as expected from Fig.\ \ref{fig:kmaxMono}, but simulations in the whole range $\phi_c \in [0.1,1]$ show similar results.  As expected, the maximum average occupation number is $\left < n^\pm \right >\approx 0.5$. Making use of Eqn.\  \eqref{eq:asym} and  Fig.\ \ref{fig:kmaxMono}, for axion monodromy inflation, the total helicity asymmetry in the number density of left- versus right-helicity fermions is given approximately by
\begin{align}
\Delta n_{\rm max} \approx 0.5 \left ( {C\over f} \right )^3 \, ,
\label{eq:maxAsymMono}
\end{align}
where the subscript refers to the maximum asymmetry. The value of $m^2 / (C/f)$ for which the maximum asymmetry occurs is a model-dependent quantity, but for the model of axion monodromy inflation studied here $m^2 / (C/f)\sim 0.05$. 
The result of Eqn.\ \eqref{eq:maxAsymMono} is valid, with good accuracy, for  $0.1 \le \phi_c \le 1$. The helicity asymmetry in the monodromy case behaves in a very similar fashion to that in the $m^2\phi^2$ case, however, with a different magnitude and a different value of the ratio $m^2 /(C/f)$ for which the maximum asymmetry is achieved. Both the exact magnitude of the maximum asymmetry and the corresponding value of $m^2 /(C/f)$ depend on $\phi_c$.

\section{Conclusion}

In this work we have studied the behavior of derivatively-coupled, massive fermionic degrees of freedom during, and immediately following axion-driven inflation. When the masses of the fermions are degenerate, these fermions pair up to make Dirac fermions and the coupling is to the axial-vector current.

During inflation,  the motion of the inflaton is monotonic and the coupling of the fermions to the rolling axion acts as an effective chemical-potential for helicity.  This chemical potential biases the gravitational production of helicity states of the fermions. In the absence of the coupling, massive fermions are produced gravitationally by the expansion of spacetime equally in all helicity states. This production is most efficient for light fermions; the production rate is exponentially suppressed by the ratio of the fermion mass to  the Hubble rate. When the axion coupling is turned on, we have shown that one helicity state is produced more efficiently while the other is strongly suppressed. This mechanism allows for the gravitational production of heavy fermion states that would otherwise be heavily suppressed.

Following inflation, as the axion oscillates, the effective frequency of the fermion field varies non-adiabatically resulting in particle production. These oscillations also mean the axion velocity changes sign, resulting in the production of both fermion  helicities. The damping of the inflaton amplitude after the end of inflation, due to Hubble friction, results in biased fermion production in favor of the helicity state that is excited during inflation. This is due to the declining efficiency of the production rate as the axion damps combined with the fact that production of the second helicity does not begin until the axion velocity changes sign. For certain combination of the fermion mass and axion-fermion coupling, the produced fermions can have a significant helicity asymmetry.  This helicity asymmetry is robust to the details of the reheating history, at least in the limited models we considered here, because it requires the amplitude of the axion oscillations to decay faster than $1/a(t)$, which does not need any significant tuning of the model.

We studied the post-inflationary evolution of the particle number in both quadratic chaotic inflation -- $m_\phi^2 \phi^2$ inflation -- and in the simplest model of axion monodromy inflation. While there are minor numerical differences between the models, broadly we find very similar behavior. In both cases the average particle number is solely defined by the combination $m^2 / (C/f)$ in the regime of large coupling, while there is increasing deviation from this behavior as we decrease the coupling. Further,  in both models the range of excited wavenumber shrinks with time as well as the production efficiency of any single event, leading to the first production events giving the dominant contribution to the final particle spectrum. The average particle number remains almost constant after the second production event for $C/f \gg 1$. Finally, the maximum achievable helicity asymmetry in the number density of produced fermions, $\Delta n$, scales as $(C/f)^3$, with a proportionality factor of ${\cal O}(1)$, which depends on the specific inflationary model.

Neutrino helicity is equal to lepton number in the standard model of particle physics, therefore the asymmetric production of helicity may be important for the generation of the matter-antimatter asymmetry in the Universe. We apply the results of this work in a companion paper \cite{Adshead:2015jza}, where we propose that an axion-inflaton coupled to left-handed standard-model neutrinos as a  mechanism for the observed baryon asymmetry in the Universe via leptogenesis.


\acknowledgments
We thank  Rob Leigh, Tom Faulkner and especially Jessie Shelton for useful discussions. EIS gratefully acknowledges support from a Fortner Fellowship at the University of Illinois at Urbana-Champaign. PA gratefully acknowledges the hospitality  of the Aspen Center for Physics and support by National Science Foundation Grant No. PHYS-1066293.

\appendix

\section{Some 2-component spinor notations}\label{app:notation}

In this Appendix, we briefly review the 2-component spinor notations and conventions we have used in this paper. Our notations follow \cite{Dreiner:2008tw}, to which we refer the reader for more detail.  We will use un-dotted indices to denote fields transforming under the left-handed representation of the Lorentz group, $(\frac{1}{2}, 0)$ ($\psi_{\alpha}\to M_{\alpha}{}^{\beta}\psi_{\beta}$), and dotted indices to denote fields that transform under the right-handed representation of the Lorentz group, $(0, \frac{1}{2})$, ($\psi^{\dagger}_{\dot\alpha}\to (M^*)_{\dot\alpha}{}^{\dot\beta}\psi^\dagger_{\dot\beta}$). There are two additional irreducible representations of the Lorentz group which are equivalent to the left- and right-handed representations above. The fields that transform under these representations are denoted by raised indices, $\psi^\alpha$ and $\psi^{\dot\alpha}$.  It is convenient to consider $\psi^\alpha$ as a row vector, and $\psi_\alpha$ as a column vector, while $\psi^\dagger{}^{\dot\alpha} = (\psi^\alpha)^{\dagger}$ as a column vector and $\psi^\dagger_{\dot\alpha}$ is a row vector. We will use the summation convention that repeated pairs of upper and lower indices are summed.

These spinor indices are raised and lowered using the spinor metric tensors, denoted by the 2D antisymmetric  $\epsilon$ symbol, where our conventions will be
\begin{align}
\epsilon^{12} = -\epsilon^{21} = \epsilon_{21} = -\epsilon_{12} = 1,
\end{align}
and we formally define $\epsilon_{\dot\alpha\dot\beta} = (\epsilon_{\alpha\beta})^*$ and $\epsilon^{\dot\alpha\dot\beta} = (\epsilon^{\alpha\beta})^*$. Note that these conventions imply
\begin{align}
\epsilon^{\gamma\delta} = -\epsilon^{\gamma\alpha}\epsilon^{\delta\beta}\epsilon_{\alpha\beta}, \quad \epsilon_{\gamma\delta} = -\epsilon_{\gamma\alpha}\epsilon_{\delta\beta}\epsilon^{\alpha\beta}.
\end{align}
These epsilon symbols satisfy
\begin{align}
\epsilon_{\alpha\beta}\epsilon^{\gamma\delta} = -\delta^\gamma_{\alpha}\delta^{\delta}_{\beta}+\delta^\delta_{\alpha}\delta^{\gamma}_{\beta}, \quad \epsilon_{\dot\alpha\dot\beta}\epsilon^{\dot\gamma\dot\delta} = -\delta^{\dot\gamma}_{\dot\alpha}\delta^{\dot\delta}_{\dot\beta}+\delta^{\dot\delta}_{\dot\alpha}\delta^{\dot\gamma}_{\dot\beta},
\end{align}
and
\begin{align}
\epsilon_{\alpha\beta}\epsilon^{\beta\gamma} = \delta^{\gamma}_{\alpha}, \quad \epsilon_{\dot\alpha\dot\beta}\epsilon^{\dot\beta\dot\gamma} = \delta^{\dot\gamma}_{\dot\alpha}.
\end{align}
We use  the sigma, or Pauli matrices, $\sigma^{a}_{\alpha\dot\beta}$ and $\bar{\sigma}^{a \dot\alpha\beta}$, defined by
\begin{align}\label{eqn:sigdef}
\sigma^0 = \bar{\sigma}^0 = \(\begin{array}{cc} 1 & 0 \\ 0  & 1 \end{array}\), \quad  \sigma^1 = -\bar{\sigma}^1 = \(\begin{array}{cc} 0 & 1 \\ 1  & 0 \end{array}\),\\
\sigma^2 = -\bar{\sigma}^2 = \(\begin{array}{cc} 0 &- i \\ i  & 0 \end{array}\), \quad  \sigma^3 = -\bar{\sigma}^3 = \(\begin{array}{cc} 1 & 0 \\ 0  & -1 \end{array}\).
\end{align}
These sigma matrices are hermitian, and are defined with a contra-variant (upper) spacetime index. Denoting the 3-vector of Pauli matrices $\vec{\sigma} \equiv (\sigma^1, \sigma^2, \sigma^3) $, the Eqn.\ \eqref{eqn:sigdef} is equivalent to
\begin{align}
\sigma^{\mu} = (1\!\!1_{2 \times2}, \vec{\sigma}), \quad \bar{\sigma}^{\mu} = (1\!\!1_{2 \times2}, -\vec{\sigma}).
\end{align} 

For convenience, we work only with left-handed fields, that is, fields that transform under the $(\frac{1}{2}, 0 )$ representation of the Lorentz group. Since these representations are related by hermitian conjugation, right-handed spinors are then simply conjugates of left-handed spinors. 

\bibliographystyle{JHEP}
\bibliography{FermiReheat}

\end{document}